\documentclass[superscriptaddress,pre,floatfix,twocolumn,amsmath,amssymb,aps,showkeys]{revtex4-1}

\usepackage{graphicx}
\usepackage{amsmath}
\usepackage{amssymb}
\usepackage{dcolumn}
\usepackage{color}
\usepackage{multirow}
\usepackage{bm}

\usepackage{subfigure}
\usepackage{tabularx}
\usepackage{booktabs}
\begin{document}
\title{Complex phase diagram and supercritical matter}
\author{Xiao-Yu Ouyang}
\author{Qi-Jun Ye}
\email{qjye@pku.edu.cn}
\affiliation{State Key Laboratory for Artificial Microstructure and Mesoscopic Physics, Frontier Science Center for Nano-optoelectronics and School of Physics, Peking University, Beijing 100871, P. R. China}
\author{Xin-Zheng Li}
\email{xzli@pku.edu.cn}
\affiliation{State Key Laboratory for Artificial Microstructure and Mesoscopic Physics, Frontier Science Center for Nano-optoelectronics and School of Physics, Peking University, Beijing 100871, P. R. China}
\affiliation{Interdisciplinary Institute of Light-Element Quantum Materials, Research Center for Light-Element Advanced Materials, and Collaborative Innovation Center of Quantum Matter, Peking University, Beijing 100871, P. R. China}
\affiliation{Peking University Yangtze Delta Institute of Optoelectronics, Nantong, Jiangsu 226010, P. R. China}
\date{\today}


\begin{abstract}
    The supercritical region is often described as uniform with no definite transitions.
    The distinct behaviors of the matter therein (as liquid-like and gas-like), however, suggest ``supercritical boundaries".
    Here, we provide a mathematical description of these phenomena by revisiting the Lee-Yang (LY) theory and introducing a complex phase diagram, i.e. a 4-D one with complex $T$ and $p$.
    While the traditional 2-D phase diagram with real $T$ and $p$ values (the physical plane) lacks LY zeros beyond the critical point, preventing the occurrence of criticality, the off-plane zeros in this 4-D scenario possess critical anomalies in various physical properties.
    For example, when the isobaric heat capacity $C_p$, which is a response function of the system to $T$, is used to separate the supercritical region, this 4D complex phase diagram can be visualized by reducing to a 3D one with complex $T$ and real $p$.
    Then, we find that the supercritical boundary defined by $C_p$ shows perfect correspondence with the projection of the edges of the LY zeros with complex $T$ in this 3D phase diagram on the physical plane, whilst in conventional LY theory these off-plane zeros are neglected.
    The same relation applies to the isothermal compression coefficient $K_T$ (or $\kappa_T$) which is a response function of the system to $p$, where complex $p$ should be used.
    This correlation between the Widom line and the edges of LY zeros is demonstrated in three systems, i.e., van der Waals model, 2D Ising model and water, which unambiguously reveals the incipient phase transition nature of the supercritical matter.
    With this extension of the LY theory and the associated new findings, a unified picture of phase and phase transition valid for both the phase transition and supercritical regions is provided, which should  apply to the complex phase diagram of other thermodynamic state functions.
\end{abstract}

\maketitle
\section{Introduction}
The supercritical behavior in real systems plays a crucial role in both fundamental research and emerging applications.
For example, supercritical water not only subtly shapes our planet but also serves as an ecologically benign solvent in chemical reactions and waste management~\cite{WeingärtnerFranck2005,Savage1999,Keppler1996}.
Starting from Cagniard de la Tour, studies of supercritical matter had greatly expanded our comprehension of states of matter in the last $\sim$200 years~\cite{Cagniard1822,Andrews1869,vanderWaals1873,Postorino_Tromp_Ricci_Soper_Neilson_1993,Smith_Kay_1999, McMillan_Stanley_2010,Galli_Pan_2013,DSouza_Nagler_2015,Cheng_Mazzola_Pickard_Ceriotti_2020,Cockrell_Trachenko_2022,Berche2009}.
To illustrate the continuity of the gaseous and liquid states in the supercritical region, Andrews established the concepts of critical temperature ($T$) and pressure ($p$)~\cite{Andrews1869}.
Van der Waals (vdW) continued this topic by revealing the equation of state (EOS) in his real gas model, which inherently suggests the lack of phase transition beyond the critical point~\cite{vanderWaals1873}.
Since then, the supercritical matter was usually introduced as a single phase~\cite{williamsSupercriticalFluidMethods2000,kiranSupercriticalFluids2000,Proctor2020}.

However, the emergence of ``supercritical boundaries'' challenges this conventional understanding~\cite{Frenkel1947,xuRelationWidomLine2005,simeoniWidomLineCrossover2010,brazhkinTwoLiquidStates2012,trachenkoCollectiveModesThermodynamics2015,ahnEnhancementServiceLife2020}.
By examining the maxima of the isobaric heat capacity $C_p$, one can separate the supercritical region, e.g. as liquid-like and gas-like subregions beyond the vaporization critical point~\cite{simeoniWidomLineCrossover2010,galloWidomLineDynamical2014,luoBehaviorWidomLine2014}.
Not surprisingly, boundaries defined by dynamics such as transverse oscillations of the particles are also spotted~\cite{chenObservationFragiletostrongDynamic2006,brazhkinTwoLiquidStates2012,trachenkoCollectiveModesThermodynamics2015,ahnEnhancementServiceLife2020,wangDirectLinksDynamical2017,lupiDynamicalCrossoverIts2021}.
While crossover phenomena serves as alternatives to phase transition, different boundaries represented by the Widom line (the line of maximum correlation length), the Frenkel line (the line where the oscillatory motion ceases), and the Fisher-Widom line (the line where the oscillatory decay in the radial distribution function becomes present) emerge~\cite{xuRelationWidomLine2005,Frenkel1947,brazhkinTwoLiquidStates2012,fisherDecayCorrelationsLinear1969}, contrasting the single boundary in the phase transition region which encapsulates all critical behaviors.
Notably, even the definition using Widom line leads to different evaluations by the maxima of $C_p$ and the isothermal compression coefficient $K_T$, respectively.
This disparity calls for delving deeper into the structure of supercritical region, aiming to interpret the macroscopically invisible phase boundaries, multiple supercritical boundaries, and crossover phenomena in this region~\cite{gartnerManifestationsMetastableCriticality2021,palmerMetastableLiquidLiquid2014,poolePhaseBehaviourMetastable1992}.
%

%
To do this, it's imperative to reveal its mathematical structures.
Similar achievements have been reached by Lee and Yang in interpreting phase transition and criticality.
In two milestone articles~\cite{yangStatisticalTheoryEquations1952,leeStatisticalTheoryEquations1952}, they found that the behaviors of zeros of the grand partition function, i.e., Lee-Yang (LY) zeros for the complex external magnetic field (chemical potential), determine the phase transition of the 2D Ising model (2D lattice gas).
The non-analytical changes in state functions of the system such as magnetization happen only when complex LY zeros fall onto the real axis in the thermodynamic limit.
Fisher generalized LY's theory to the canonical ensemble and defined Fisher zeros for the complex temperature ($T$)~\cite{fisherLecturesTheoreticalPhysics1965}.
Now, it is customary to analyze the phase transition phenomena using such LY or Fisher zeros, with applications extending to the studies of
their experimental measurements~\cite{pengExperimentalObservationLeeYang2015,weiLeeYangZerosCritical2012}, nonequilibrium problems~\cite{brandnerExperimentalDeterminationDynamical2017,flindtTrajectoryPhaseTransitions2013}, protocols of quantum
simulators~\cite{francisManybodyThermodynamicsQuantum2021,krishnanMeasuringComplexpartitionfunctionZeros2019,xuProbingFullDistribution2019,gnatenkoTwotimeCorrelationFunctions2017},
and dynamical quantum phase transitions~\cite{heylDynamicalQuantumPhase2018,heylDynamicalQuantumPhase2013}.
Recent studies of quantum chromodynamics (QCD) models demonstrate a correlation between the LY zeros and the crossover behavior~\cite{basarUniversalityLeeYangSingularities2021,connellyUniversalLocationYangLee2020}.
Heuristically, one can understand the supercritical boundaries by establishing theoretical and numerical connections between the supercritical behaviors and the complex LY zeros in realistic condensed matter systems.
Without losing generality, we use two typical thermodynamic state functions (which can also be viewed as fields, as will be explained later) $T$ and $p$ and consider zeros in the $T$-$p$ phase diagram.
A complex space $\tilde{x}$ of $T\to \tilde{T} = T + i\tau$ and $p\to \tilde{p}= p + i\zeta$ is employed.
By analytic continuation, the partition function can be represented in terms of complex zeros corresponding to $\tilde{T}$- or $\tilde{p}$- perspectives, as
\begin{equation}
    \begin{aligned}
        Z(\tilde{T},\tilde{p}) & = Z_{\tilde{p}}(\tilde{T}) = e^{g_{\tilde{p}}(\tilde{T})} \prod_{k=1}^{\infty} \left( 1-\tilde{T}/{{\tilde{T}}^*_{\tilde{p},k}} \right)   \\
                               & =  Z_{\tilde{T}}(\tilde{p}) = e^{g_{\tilde{T}}(\tilde{p})} \prod_{l=1}^{\infty} \left( 1-\tilde{p}/{{\tilde{p}}^*_{\tilde{T},l}} \right), \\
    \end{aligned}
    \label{partition function}
\end{equation}
where $\tilde{T}^*_{\tilde{p},k}$ is the $k$-th non-zero root for $Z_{\tilde{p}}(\tilde{T})=0$ at given $\tilde{p}$, and $\tilde{p}^*_{\tilde{T},l}$ is defined similarly, please see Appendices~\ref{app: factorization} and \ref{app: multiple field} for details.
For an ordinary $T$-$p$ phase diagram, the physical plane consists of the real axes of $T$ and $p$.
Here, due to the dependency of $\tilde{T}^*_{\tilde{p},k}$ on $\tilde{p}$ and $\tilde{p}^*_{\tilde{T},l}$ on $T$,
$\tilde{T}^*_{\tilde{p}}$ and $\tilde{p}^*_{\tilde{T}}$ manifest a unified cluster of zeros in a 4-D complex space $\mathcal{C}^2$.
When taking physical values of $T$ ($p$), i.e., their real values, $\tilde{p}^*_{\tilde{T}=T}$ ($\tilde{T}^*_{\tilde{p}=p}$) returns to LY (Fisher) zeros.
Acknowledging that zeros in the physical plane (real $T$ and $p$) locate phase boundaries and critical points, we emphasize here that zeros outside the physical plane are
of crucial importance and responsible for the anomalies in the supercritical region.
Using this 4D complex phase diagram, we investigate the supercritical phenomena.
Different from the traditional scenario, the high-dimensional space with extra imaginary axes allows for a comprehensive description of the physical properties.
We found the extreme line for each response function exactly corresponds to its closest LY zeros to the physical plane, i.e. the LY edges.
For the simplest cases, the maxima of the isobaric heat capacity $C_p$ (isothermal compression coefficient $K_T$) as the response function
to changed $T$ ($p$), show close coincidence with the $T$- ($p$-) edges of the same cluster of zeros.
Through a 3-D complex phase diagram with one of the imaginary axes contracted, visible insights into the supercritical behavior can be depicted.
Distinct zeros dominate on each side of the edge, resulting in different properties on the two sides of the supercritical region.
These zeros present good illustration of phase transitions and crossovers, which offers us an intuitive tool to consolidate critical and supercritical matter.
This paper is organized as follows.
In Sec.~II, we explain the methods we used for calculating the thermodynamic properties and LY zeros in the vdW model, 2D Ising model, and water.
How the complex phase diagram and LY zeros in it can be used to describe the thermodynamic properties of the supercritical region is explained in Sec.~III.
Then, we provide extensive discussions on the meaning and implications of this concept in Sec.~IV.
The conclusion is drawn in Sec.~V.
For the reader's convenience, the paper is also supplemented with three appendices which clarify some finer theoretical and technical details.
With such an arrangement, we hope a complete explanation of the theoretical and computational findings and their implications can be conveyed by us to a wide
range of audience, so that a route toward a unified picture of phase and phase transition for both the phase transition and the supercritical regions appears
clear, within the framework of LY theory.

\section{Methods}
To demonstrate the universal nature of the complex phase diagram and its connection to supercritical behaviors, our analysis proceeded along two fronts.
Firstly, we consider model systems, represented by the vdW and 2D Ising models, where the partition function and LY zeros are analytically tractable.
Secondly, we tackle more realistic systems, exemplified by water with TIP4P interactions, where the LY zeros are evaluated from molecular dynamic (MD) simulations.
We adopted the Widom lines as the representative supercritical boundaries, since their evaluations only depend on basic thermodynamic properties, such as volume and enthalpy.
Specifically, we conduct calculations of their thermodynamic properties and estimate the Widom line by extreme lines of $C_p$ and $K_T$ (or $\kappa_T$) for each system.

\subsection{vdW model}

\subsubsection{Thermal properties and extreme lines}
The vdW model is the simplest real gas model where particles interact and occupy finite volumes.
Its equation of state is written as
\begin{equation}
    \left[\frac{p}{p_{{c}}}+\frac{3}{\left(V / V_{{c}}\right)^2}\right]\left(\frac{V}{V_{{c}}}-\frac{1}{3}\right)=\frac{8 T}{3 T_{{c}}},
\end{equation}
where $(p_c, V_c, T_c)$ is the critical point of vdW fluid.
We take it to be $(1,1,1)$ for convenience.
In order to derive thermodynamic properties and determine supercritical boundaries, we start from analytical formulas of free energy and its derivatives.
The Gibbs free energy is given by
\begin{equation}
    \begin{split}
        G(T, &p)=\\
        &-T\left[C+\frac{3}{2} \ln T+\ln \left(V-\frac{1}{3}\right)+\frac{9}{8 T V}-\frac{3 p V}{8 T}\right].
    \end{split}
    \label{Eq:freeEnergy}
\end{equation}
The enthalpy $H$ is
\begin{equation}
    \begin{split}
        H &=G+TS = G - T\left(\frac{\partial G}{\partial T}\right)_p \\
        &=  -T^2\left[\frac{\partial }{\partial T}\left(\frac{G}{T}\right)\right]_p=\frac{3}{2} T-\frac{9}{8 V}+\frac{3 p V}{8}.
    \end{split}
\end{equation}
The isobaric heat capacity reads
\begin{equation}
    C_{p} =\left(\frac{\partial H}{\partial T}\right)_p=\frac{3}{2}+\frac{4TV^3}{4TV^3-(3V-1)^2}.
    \label{Eq:C_p}
\end{equation}
The volume $V$ is the partial derivative of $G$ to pressure $p$, as
\begin{equation}
    V=\left(\frac{\partial G}{\partial p}\right)_T,
\end{equation}
whose expression as a function of $(T, p)$ will be given later, as in Eq.~(\ref{Vsolution}).
And the isothermal compression coefficient is given by
\begin{equation}
    K_T=-\left(\frac{\partial V}{\partial p} \right)_T=\frac{(3V-1)^2V}{6\left[4TV^3-(3V-1)^2 \right]}.
    \label{Eq:K_T}
\end{equation}

\begin{figure}
    \centering
    \includegraphics[width=1\linewidth]{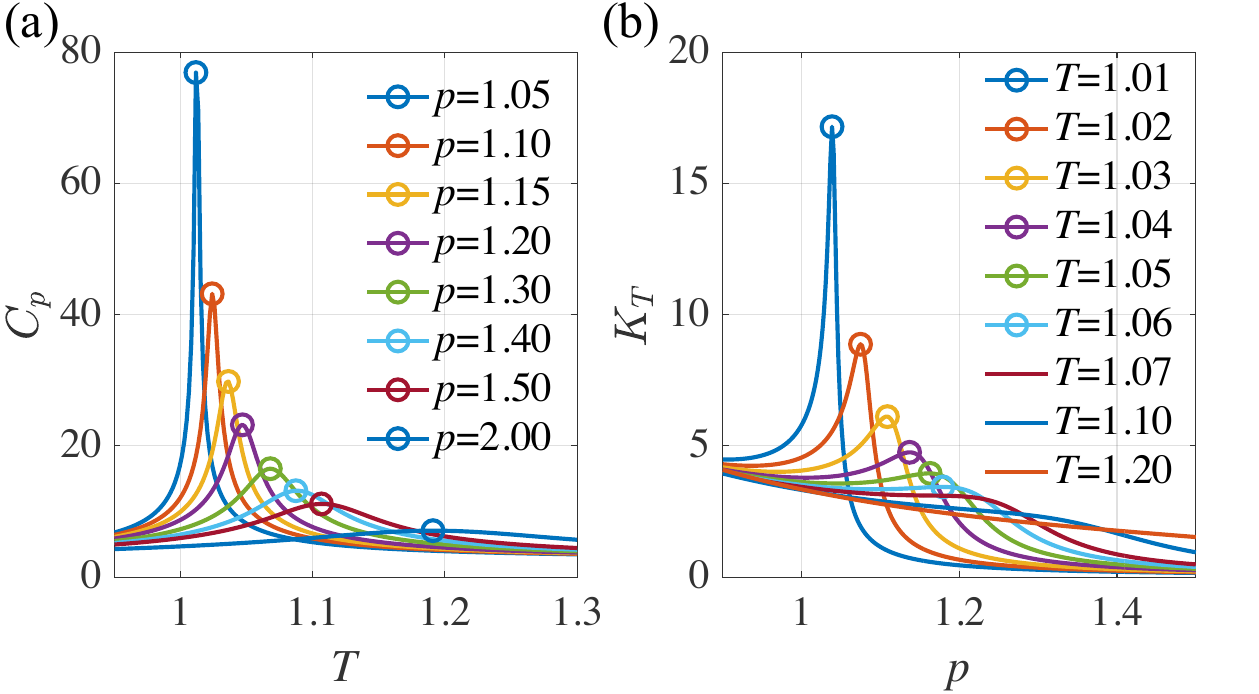}
    \caption{The curves of isobaric heat capacity $C_p$ and isothermal compression coefficient $K_T$ for the vdW model in the supercritical region.
    (a) $C_p$ along isobars, with the peaks of $C_p$ marked with circles.
    (b) $K_T$ along isotherms, with the peaks of $K_T$ marked with circles. The extreme value vanishes when $T=1.07$ or higher.}
    \label{fig:vdwWidom}
\end{figure}

With those formulas, we can analytically locate $C_p$ and $K_T$ extreme lines along isotherms and isobars, respectively.
This is done by solving $\partial C_p/\partial T=0$ and $\partial K_T /\partial p=0$.
For the other kinds of extreme lines, please see analytic expressions in Ref.~\cite{brazhkinVanWaalsSupercritical2011}.
Below the critical point, the two extreme lines converge to a single coexistence line, manifesting the phase boundary of gas and liquid.
This can also be calculated through Maxwell's construction~\cite{johnstonThermodynamicPropertiesVan2014}.
%
%
The results for supercritical region are shown in Fig.~\ref{fig:vdwWidom}.
We note that $K_T$ extreme lines vanish at $T\sim 1.07$ (see Fig.~(\ref{fig:vdwWidom})(b)).
It implies the competition between the LY zeros terms and the other analytic terms, please see discussions later in Appendix~\ref{App: termination of extreme line}.

\subsubsection{Density of zeros}
Following the idea of Lee and Yang~\cite{yangStatisticalTheoryEquations1952,leeStatisticalTheoryEquations1952}, we shall analytically extend the domain
of the Gibbs free energy function to the complex space, as $G(T,p)\to G(\tilde{T},\tilde{p})$, where the tilde is for complex variables.
We note that the zeros of partition function are equivalent to the singularities of the Gibbs free energy.
From Eq.~(\ref{Eq:freeEnergy}), we see that the singularities of $G$ come from three cases: i) singularities of $V(\tilde{T},\tilde{p})$; ii) $T=0$; iii) $V=1/3~\text{or}~0$.
The volume $V$ satisfies the cubic equation
\begin{equation}\label{veq}
    V^3-\left(\frac{1}{3}+\frac{8 T}{3 p}\right) V^2+\frac{3}{p} V-\frac{1}{p}=0.
\end{equation}
The latter two cases for the singularities of $G$ correspond to trivial zeros at $T=0$ or $p=\infty$ according to Eq.~(\ref{veq}).
Therefore, we only consider the first case, i.e., singularities of $V(\tilde{T},\tilde{p})$.
The analytical expression of the complex Gibbs free energy in Eq.~(\ref{Eq:freeEnergy}) requires $V(\tilde{T},\tilde{p})$, which can be obtained from Eq.~(\ref{veq}).
It has two complex roots and one real root.
We retain the latter one since it is the only physical solution, reading as
\begin{equation}
    V=A+ \sqrt[3]{Q+\sqrt{D}}+\sqrt[3]{Q-\sqrt{D}},
    \label{Vsolution}
\end{equation}
where
\begin{equation}
    \begin{aligned}
        A=   & ~ \frac{1}{9}\left(1+\frac{8 T}{p}\right),                                           \\
        Q=   & ~ A^3-\frac{3 A}{2 p}+\frac{1}{2p},                                                  \\
        D  = & ~ \frac{A^3}{p}-\frac{3 A^2}{4 p^2}-\frac{3 A}{2 p^2}+\frac{1}{p^3}+\frac{1}{4 p^2}.
    \end{aligned}
\end{equation}
Without losing generality and for convenience, we choose the branch of the square function to be $\sqrt{z}=\sqrt{|z|}\exp{(i\mathrm{Arg}[z]/2)}$, where $\mathrm{Arg}[z]\in [-\pi/2,3\pi/2)$.
For the cubic function, we always adopt its real root, i.e., when $z=a$ ($a \in \mathbb{R}$ and $a>0$), $\sqrt[3]{z}=\sqrt[3]{a}$; when $z=-a$, $\sqrt[3]{z}=-\sqrt[3]{a}$.
Accordingly, we perform the continuation as
\begin{equation}\label{veq4}
    \sqrt[3]{z}= \begin{cases}\sqrt[3]{|z|}\exp{(i\mathrm{Arg}[z]/3)}, & \Re[z]\geq 0 \\ \sqrt[3]{|z|}\exp{\{i(\mathrm{Arg}[z]+2\pi)/3\}}, & \Re{[z]}<0\end{cases}
\end{equation}
With this choice of branches, the cubic function undergoes a discontinuity if the complex path of $z$ crosses the imaginary axis.
Hence, the loci of $V$'s --- as well as $G$'s --- singularity, can be written as
\begin{equation}
    \begin{aligned}
        \left\{(\tilde{T},\tilde{p})~\Bigg| ~\Re\left[Q(\tilde{T},\tilde{p})+\sqrt{D(\tilde{T},\tilde{p})} \right] =0 \text{ or } \right. \\
        \left.
        \Re\left[Q(\tilde{T},\tilde{p})-\sqrt{D(\tilde{T},\tilde{p})}\right]=0  \right\}.
    \end{aligned}
    \label{vdWsingular}
\end{equation}
In principle, LY zeros can be located according to Eq.~(\ref{vdWsingular}).

\begin{figure}[htbp]
    \centering
    \includegraphics[width=0.95\linewidth]{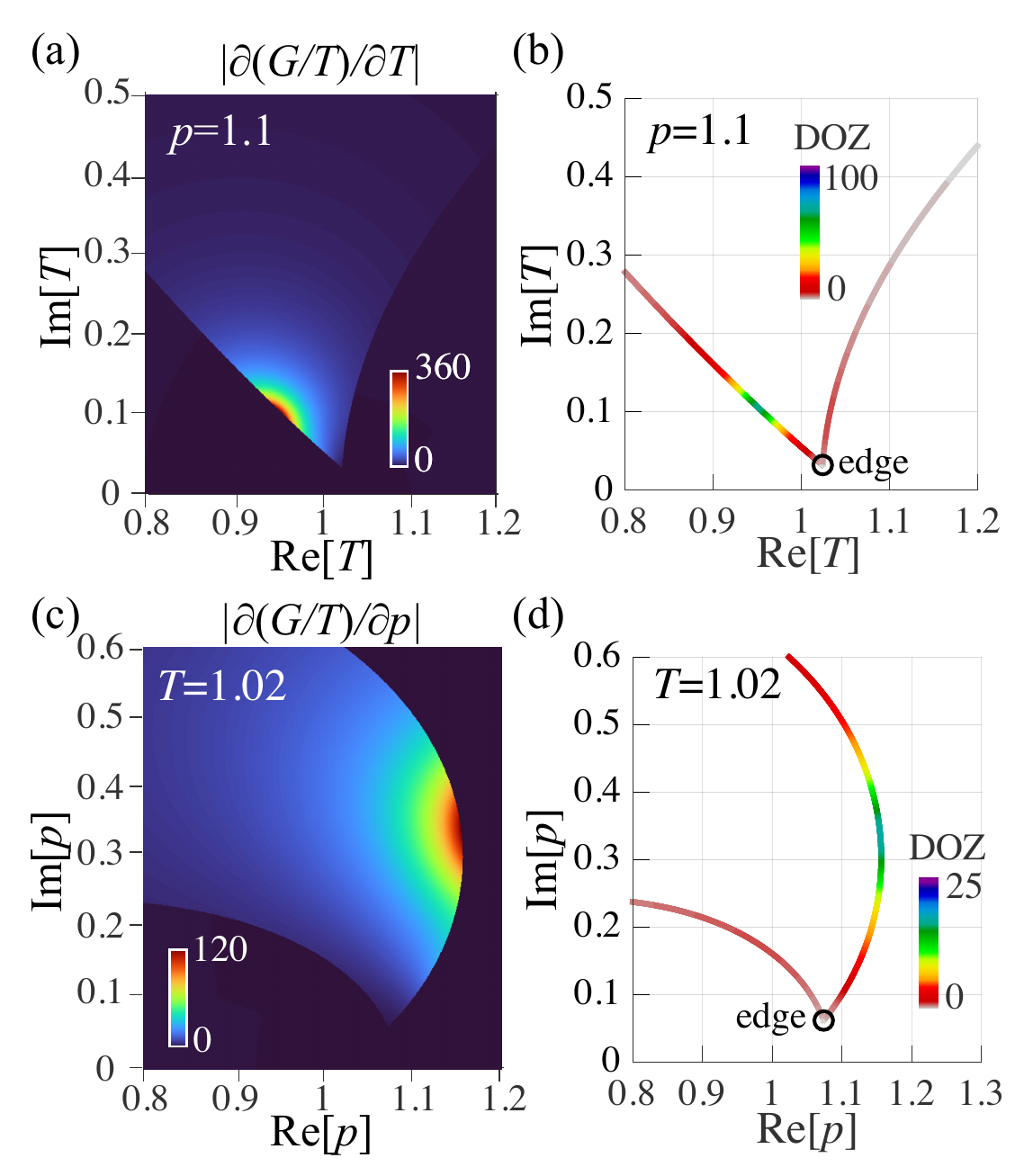}
    \caption{
    The derivatives' singularity of complex free energy term $G/T$ and LY zeros for the vdW fluid, calculated using Eq.~(\ref{Eq:freeEnergy}) and (\ref{Eq:vdwDensity}).
    (a) The modulus of the derivative of $G/T$ to temperature $T$, plotted on the complex $T$ plane when the pressure $p=1.1$.
    (b) The loci and density distribution of $T$-zeros under the same conditions of (a).
    (c) The modulus of the derivative of $G/T$ to pressure $p$, plotted on the complex $p$ plane when the pressure $T=1.02$.
    (d) The loci and density distribution of $p$-zeros under the same conditions of (c).
    For visual convenience, the symmetrized lower half complex planes are now shown.
}
    \label{fig:vdwFreeEnergy}
\end{figure}
In this manuscript, we adopt a different path from directly using Eq.~(\ref{vdWsingular}).
But we note that they are completely equivalent.
This is enabled by putting Eqs.~(\ref{Vsolution})-(\ref{veq4}) to Eq.~(\ref{Eq:freeEnergy}), and monitoring the discontinuity of the Gibbs free energy.
According to Ref.~[\onlinecite{leeStatisticalTheoryEquations1952}], similar to the electrostatic potentials, continuous and linear-distributed zeros
is the source of the discontinuity of the derivative of the Gibbs free energy on the complex plane.
Therefore, using the analogy of the Gaussian theorem, the density of LY zeros $\rho(\tilde{x})$ satisfies
\begin{equation}
    \left[\frac{\partial (G/T)}{\partial \widetilde{x}}\right]_{\widetilde{x}_{0+}}-\left[\frac{\partial (G/T)}{\partial \widetilde{x}}\right]_{\widetilde{x}_{0-}}=e^{i \alpha}\cdot 2 \pi \rho\left(\widetilde{x}_0\right),
    \label{Eq:vdwDensity}
\end{equation}
where $\alpha$ is the angle between the zero line and the imaginary axis.
When the derivative $\partial (G/T)/\partial \widetilde{x}$ is continuous at $\tilde{x}_0$, $\rho(x_0) =0$, otherwise one get finite magnitude, $\rho(x_0)\neq 0$.
The results are shown in Fig.~\ref{fig:vdwFreeEnergy}.

\subsection{2D Ising model}
We use the square 2D Ising model with ferromagnetic interaction $J>0$, which was the model Lee and Yang employed in the initial paper of LY zeros \cite{leeStatisticalTheoryEquations1952,yangStatisticalTheoryEquations1952}.
Here, we briefly summarize the results since its zeros has been well studied previously~ \cite{degerLeeYangTheoryHigh2020,kortmanDensityZerosLeeYang1971,kimYangLeeZerosAntiferromagnetic2004,krasnytskaViolationLeeYangCircle2015,krasnytskaPartitionFunctionZeros2016,binekDensityZerosLeeYang1998}.

For a periodic 2D Ising lattice with $L\times L$ spins, the Hamiltonian reads
\begin{equation}
    H=-h\sum_i s_i-J \sum_{\langle i,j \rangle}s_i s_j,
\end{equation}
where $s_i=\pm1$, $h$ is the magnetic field, and $\langle i,j \rangle$ represents for neighboring interactions.
One could reduce the problem to an exact diagonalization of a screw building-up of length $L$.
The partition function is written in terms of a transfer matrix $\mathbf{T}$ and further its eigenvalues, as
\begin{equation}
Z(\beta,h)=\operatorname{Tr}\left\{\mathbf{T}^L\right\}=\sum_{j=1}^{2^L} \lambda_j^L.
    \label{Eq:isingPartition}
\end{equation}
Here, $\mathbf{T}$ is a $2^L \times 2^L$ matrix,
\begin{equation}
\mathbf{T}=[2 \sinh (2 \beta J)]^{L / 2} \mathbf{V}_3 \mathbf{V}_2 \mathbf{V}_1
\end{equation}
being the product of 3 matrices,
\begin{equation}
\mathbf{V}_1=\prod_{i=1}^L e^{\Theta \mathbf{X}_i}, \quad \mathbf{V}_2=\prod_{i=1}^L e^{\beta J \mathbf{Z}_i \mathbf{Z}_{i+1}}, \quad \mathbf{V}_3=\prod_{i=1}^L e^{\beta h \mathbf{Z}_i}
\end{equation}
where $\tanh \Theta=e^{-2 \beta J}$, and
\begin{equation}
\begin{aligned}
&\mathbf{X}_i&=&\mathbb{I} \otimes \mathbb{I} \otimes \cdots \otimes \sigma_x \otimes \cdots \otimes \mathbb{I} \otimes \mathbb{I}\\
&\mathbf{Z}_i&=&\mathbb{I} \otimes \mathbb{I} \otimes \cdots \otimes \sigma_z \otimes \cdots \otimes \mathbb{I} \otimes \mathbb{I},
\end{aligned}
\end{equation}
with $\mathbb{I}$ being the identity $2\times 2$ matrix, and $\sigma_x$, $\sigma_z$ being the Pauli matrices on position $i=1,...,L$.

We use an $8\times 8$ lattice, and take $J=1$ without losing generality.
Given the partition function $Z(\beta,h)$, we calculated the exact LY zeros by searching its $\beta$ and $h$ zeros.
Besides, the free energy can be immediately calculated as the logarithm of partition function.
Accordingly, heat capacity $C$ and susceptibility $\chi$ as the derivatives of free energy, and hence Widom lines are determined.

\subsection{TIP4P water}
As an example to treat realistic systems, here we show how to calculate supercritical properties and LY zeros by molecular dynamics (MD) simulation of water.

\subsubsection{Simulation details}
We performed molecular dynamics in a periodic cubic box containing 216 water molecules with an initial density of $\sim\rho=1~\mathrm{g/cm^3}$.
We use the TIP4P/2005 model \cite{abascalGeneralPurposeModel2005}, with the cutoff length of both the LJ potential and Coulomb potential set as 10~Å.
The long-range Coulomb interaction is treated by particle-particle particle-mesh solver (PPPM).
The simulations are run in the $NPT$ ensemble, with an integration time step 1~fs.
We control $T$ by the Nosé-Hoover thermostat \cite{hooverCanonicalDynamicsEquilibrium1985, noseUnifiedFormulationConstant1984} and $p$ by the Parrinello-Rahman barostat \cite{noseConstantPressureMolecular1983, parrinelloPolymorphicTransitionsSingle1981}, both with the damping time 200~fs.
All the simulations are run using LAMMPS \cite{thompsonLAMMPSFlexibleSimulation2022} (version 17Nov16) compiled with Intel C++ Compiler 16 and Intel MPI 5.1.

For the phase transition region, we performed the simulations for a mesh of configurations $p=50\sim100 ~\mathrm{atm}$ and $T=675\sim700~\mathrm{atm}$.
We use a typical sampling duration of 1~ns, and a finer one of 5~ns for the vicinity of the phase transition point.
For the supercritical region, we simulated for isobars with pressures $p_n=120+10n~\mathrm{atm}~(n=1,...,16)$ to obtain $C_p$.
For each $p$, the simulated $T$ ranges from 600~K to 900~K, with spacing $\Delta T = 5~\mathrm{K}$.
For each $(T,p)$, we first equilibrate the system for 0.1~ns and then execute runs of 1~ns.
To locate the $C_p$ maxima and the $T$-zeros more accurately, we execute runs of 5~ns at the vicinity of each $C_p$ peak, with finer spacing $\Delta T=1~\mathrm{K}$.

To obtain $\kappa_T$, we simulated for isotherms with temperatures $T_n=700+5n~\mathrm{K}~(n=1,...,16)$.
Since we have known the $C_p$ extreme line from the isobars, we execute runs of 2~ns only in the region nearby the extreme lines, e.g. pressure from $p=150~\mathrm{atm}$ to $p=176~\mathrm{atm}$ for isotherm $T=730~\mathrm{K}$, and $\Delta p=2~\mathrm{atm}$.
To better locate $\kappa_T$ maxima, we enact runs of 10ns in a smaller vicinity of the $\kappa_T$ peaks.

\subsubsection{Extreme lines}
\begin{figure}[b]
    \centering
    \includegraphics[width=1\linewidth]{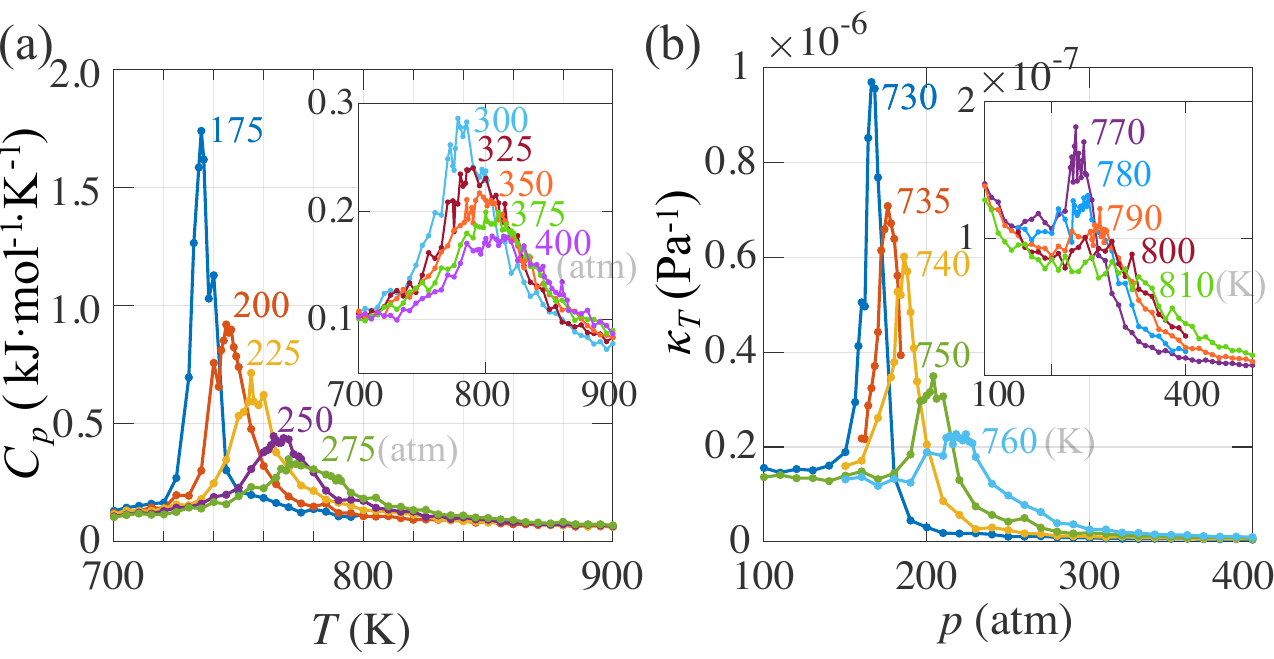}
    \caption{The simulated isothermal heat capacity $C_p$ and isobaric compressibility $\kappa_T$ of TIP4P water according to Eq.~(\ref{Eq:responseFunc}).
    (a) $C_p$ along isobars from $p=175$ to $275~\mathrm{atm}$.
    (b) A magnification for $p=300~\mathrm{atm}$ and higher.
    (c) $\kappa_T$ along isotherms from $T=730$ to $760~\mathrm{K}$.
    (d) A magnification for $T=770~\mathrm{K}$ or higher.}
    \label{fig:waterWidom}
\end{figure}
We locate the extreme lines of $C_p$ and $\kappa_T$ directly from the sampling data of enthalpy and volume.
We have
\begin{equation}
    \begin{aligned}
        C_p=\left(\frac{\partial H}{\partial T}\right)_p=\frac{\left\langle H^2\right\rangle-\langle H\rangle^2}{k_B T^2} , \\
        \kappa_T =-\frac{1}{V}\left(\frac{\partial V}{\partial p}\right)_{T}=\frac{\left\langle V^2\right\rangle-\langle V\rangle^2}{k_B T V}.
    \end{aligned}
    \label{Eq:responseFunc}
\end{equation}
Here, we use $\kappa_T$ instead of $K_T$ to characterize the response of volume to pressure, and estimate the Widom line.
This is due to the fact that $K_T$ of water decays rapidly across from liquid to gas, making it difficult to determine their peaks and hence Widom line, while $\kappa_T$ has better behavior in the supercritical region.

The results are shown in Fig.~\ref{fig:waterWidom}.
Similar to vdW case, the non-zero analytic part would take over for configurations far away from the critical point.
When $T$ is higher than 790 K, the extreme value of $\kappa_T$ becomes insignificant with a smooth crossover and limited supercritical behavior.
More details please see Appendix \ref{App: termination of extreme line}.

\subsubsection{Calculation of discrete zeros}
\label{sec: discrete zeros}

Here, we developed an approximate method for calculating LY zeros based on MD results at $(T_0, p_0)$.
This is done by estimating the density of states using the probability distribution of enthalpy $H$ and volume $V$.
By discretizing the partition function into polynomials, we could derive $p$- or $T$-zeros.
Typical extensive quantities distribute continuously, while we approximately describe them as discrete levels, e.g.,
\begin{equation}
    \rho(V)dV \to g(V\in [V_k, V_{k+1})) \Delta V = g_k \Delta V,
\end{equation}
where the volume is discretized by $N_V$ bins with bin size $\Delta V$, as $V_k = V_0 + k\Delta V$.
In the meantime, this can be derived from the partition function, via
\begin{equation}
    \begin{split}
        g_k \Delta V &= \frac{1}{Z(T,p)} \int_{V_k}^{V_{k+1}} e^{-\beta pV} \rho(T,V) dV\\
        & \approx \frac{1}{Z(T,p)} \left.\frac{-1}{\beta p}e^{-\beta pV}\right|_{V_k}^{V_{k+1}} \rho(T,(V_k+V_{k+1})/2) \Delta V\\
        & \sim \frac{1}{Z(T,p)} \frac{-1}{\beta p} e^{-\beta pV_k} \rho_k(T) \Delta V
    \end{split}
\end{equation}
Based on this, one can perform MD simulations at $(T_0,p_0)$ and extract information about $\rho_k(T_0)$ from the observed probability distribution $g_k$, via
\begin{equation}
    \rho_k(T_0) \sim g_k(T_0,p_0) e^{\beta_0 p_0 V_k}.
\end{equation}
Here, we neglect the common coefficients since only relative coefficients affect the zeros of the polynomials.
Considering the $p$-zeros at fixed $T$, the partition function can also be given in form of $g_k(T,p_0)$, as
\begin{equation}
    \begin{split}
        Z_T(p) &= \int_V e^{-\beta pV} \rho(T,V)dV \\
        &\to \sum_{k=1}^{N_V} \rho_k(T) \Delta V e^{-\beta p (V_0 + k\Delta V)}\\
        &\sim e^{-\beta p V_0}\sum_{k=1}^{N_V} g_k(T,p_0) e^{\beta p_0 V_k} \left[e^{-\beta p \Delta V}\right]^k \\
        &\sim \sum_{k=1}^{N_V} a_k \left[e^{-\beta p \Delta V}\right]^k =  \sum_{k=1}^{N_V} a_k^\prime \left[e^{-\beta (p-p_0) \Delta V}\right]^k.
    \end{split}
\end{equation}
We ignore the common coefficient $e^{-\beta p V_0}$ outside the summation in the third line since it is always nonzero throughout the complex plane.
The zeros of argument $y = e^{-\beta p \Delta V}$ can be directly solved by polynomials with the coefficients $a_k\sim g_k(T,p_0) e^{\beta p_0 V_k}$.
Or alternatively, this can also be done by solving for zeros of argument $y^{\prime} = e^{-\beta (p-p_0) \Delta V}$ with $a_k \sim g_k(T,p_0)$ and then shifting the zeros by $p_0$.
With performed MD simulations, the $T$-zeros at fixed $p$ can be calculated similarly, as
\begin{equation}\begin{split}
        Z_p(T) &\sim e^{-\beta H_0} \sum_{k=1}^{N_H} g_k(T_0,p) e^{\beta_0 H} \left[e^{-\beta \Delta H}\right]^k \\
        & \sim \sum_{k=1}^{N_V} b_k \left[e^{-\beta  \Delta H}\right]^k =  \sum_{k=1}^{N_V} b_k^\prime \left[e^{-(\beta -\beta_0) \Delta H}\right]^k,
    \end{split}
\end{equation}
where $\beta_0 = 1/(k_B T_0)$, $b_k\sim g_k(T_0,p) e^{\beta_0 H_k}$, and $b_k^{\prime} \sim g_k(T_0, p)$.

\begin{figure}[htbp]
    \centering
    \includegraphics[width=1\linewidth]{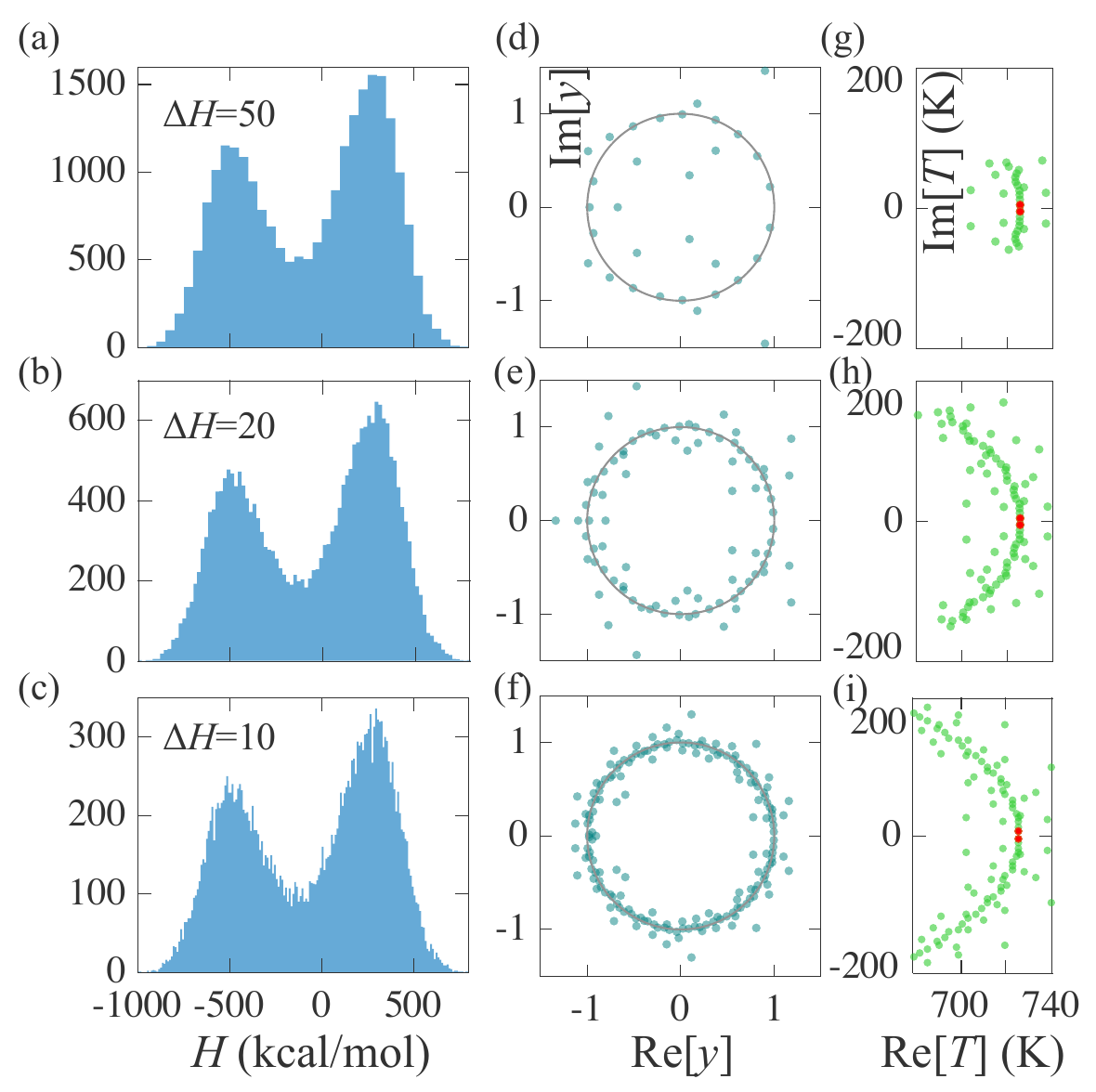}
    \caption{The histograms of enthalpy $H$ and corresponding distributions of $T$-zeros for TIP4P water, sampled at $p=160~\mathrm{atm}$ and $T=726~\mathrm{K}$.
    (a)-(c) The sampled histograms of enthalpy distribution, plotted with bin width $\Delta H=50,~20,~\text{and}~10~\mathrm{kcal/mol}$, respectively.
    (d)-(f) LY zeros expressed by zeros of $y$ at each $\Delta H$ (50, 20 and 10 kcal/mol). The unit circles are marked out for visual guidance.
    (g)-(i) LY zeros of temperature $T$ at each $\Delta H$. The LY edges are marked with red dots.}
    \label{fig:waterHist}
\end{figure}

The bin size would affect the results of $T$-zeros, as shown in Fig.~\ref{fig:waterHist}.
With the bin size $\Delta H$ varying from 50 to 10 kcal/mol~(Fig.~\ref{fig:waterHist}(a)-(c)), the zeros of $y=e^{-(\beta-\beta_0)\Delta H}$ become denser in the complex plane of $y$ (Fig.~\ref{fig:waterHist}(d)-(f)).
However, there is no significant change for the zeros of $T$ near the real axis when converting to the complex plane of $T$ (Fig.~\ref{fig:waterHist}(g)-(i)).
The more detailed distribution with a smaller bin size makes the farther $T$-zeros available for us.
However, too small bin size would bring noise of distribution, which is harmful for the accuracy.
In our calculation, we choose $\Delta H=20$ kcal/mol and $\Delta V=500~\mathrm{Å^3}$ to locate the zeros of $T$ and $p$ on the complex plane.

\section{Results}

\subsection{vdW model}

\begin{figure}
    \centering
    \includegraphics[width=0.95\linewidth]{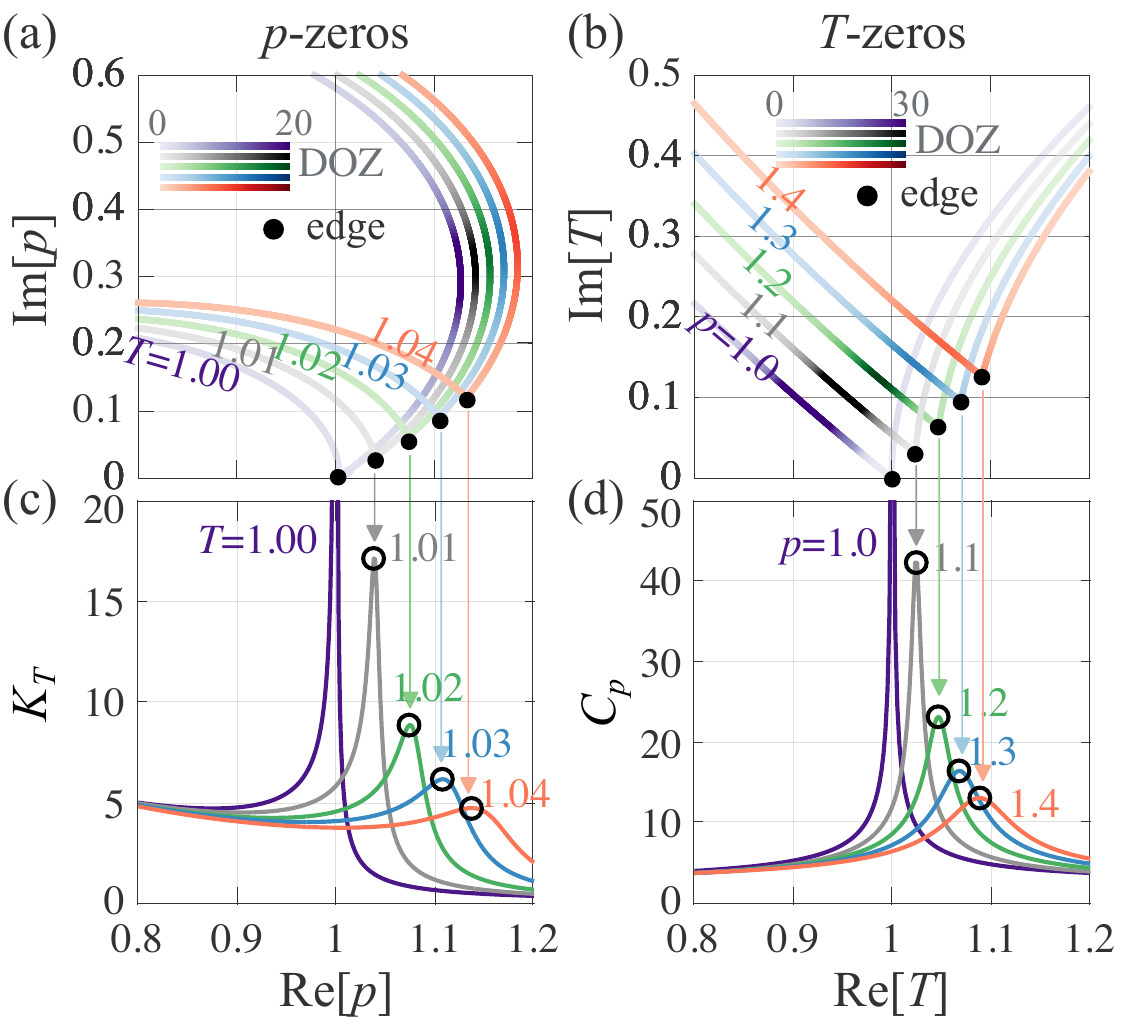}
    \caption{Lee-Yang zeros and response functions of vdW fluid in the supercritical region.
        The density of zeros (DOZ, represented by the depth of color) $\rho(\tilde{x})$ at $\mathcal{C}^2$ as 2-D slices is displayed at (a) $T=1.00,1.01,...,1.04~T_c$  and (b) $p=1.0,1.1,...,1.4~p_c$, with LY edges marked with black dots and identical lower half planes hidden.
        The associate thermal properties are also shown: (c) isothermal compression coefficient $K_T$ and (d) isobaric heat capacity $C_p$ in corresponding colors respectively, with their maxima marked in circles.
        The location where the response function $K_T$ or $C_p$ reach its maximum is very close to the projection of the corresponding LY edge with the same color on the real axis.
    }
    \label{fig:vdwZero}
\end{figure}

By utilizing the singular properties of $G(\tilde{T},\tilde{p})$ in the complex plane, we've calculated the density of zeros (DOZ) of the partition function $Z$ of vdW model.
For each $p$ or $T$, the points with nonzero DOZ converge to a line, with sharp LY edge (Fig.~\ref{fig:vdwZero}(a)(b)).
As expected beyond the critical point, the LY edges are no longer on the real axis, but move into the complex plane.
Besides this, we also noticed that the $p$-edges are intimately related to the maxima (not singularities) of $K_T$ (Fig.~\ref{fig:vdwZero}(a)(c)),
so do those of the $T$-edges and the maxima of $C_p$ (Fig.~\ref{fig:vdwZero}(b)(d)).
With $p$- ($T$-) edges become farther from the real axis, the $K_T$ ($C_p$) maxima become less sharper at almost coincident locations.

\begin{figure}[h]
    \centering
    \includegraphics[width=0.95\linewidth]{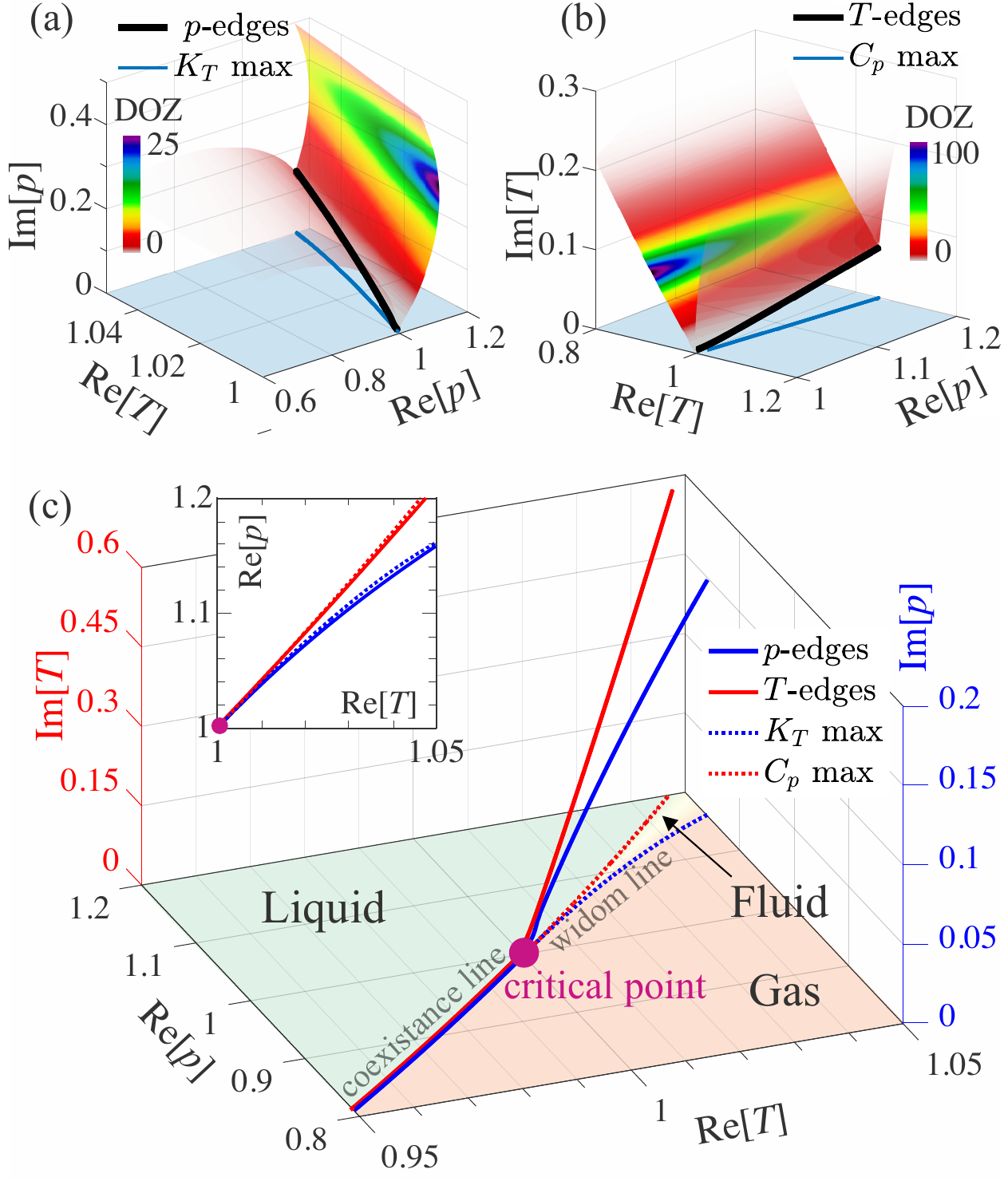}
    \caption{The complex $\tilde{T}$-$\tilde{p}$ phase diagram of the van der Waals model around the critical point.
        The 3D projection of DOZ with (a) $\mathrm{Im}[T]$ axis contracted and (b) $\mathrm{Im}[p]$ axis contracted.
        $C_p$ and $K_T$ maxima and $T$- and $p$-edges are in dark blue and black lines, respectively.
        The real plane of the phase diagram is painted blue, and identical lower half space is hidden.
        (c)
        The Lee-Yang edges corresponding to $T$ and $p$ are plotted in the 4D complex phase diagram, with an imaginary z-axis of both $T$ and $p$.
        The edges terminate at the critical point and converge to the same coexistence line in the physical plane.
        While in the supercritical region, one witnesses different edges in the complex plane and hence different extreme lines in the physical plane.
        The inset shows a close connection between the projection of edges and the Widom line, with the latter estimated by $C_p$ and $K_T$ maxima.
    }
    \label{fig:vdwCpd}
\end{figure}

To demonstrate these in details, we quantify the geometric relationship between the response functions and LY zeros using 3D-plots in Fig.~\ref{fig:vdwCpd}(a)\&(b).
The $x-y$ plane is the physical plane with real $p \& T$ and the one used in conventional phase diagrams, while the $z$ axis represents imaginary $p$ or $T$.
The $K_T$ or $C_p$ extreme line shows apparent correspondence to the line of $p$- or $T$- edges, respectively.
The correspondence arises from the nature of the response function itself: the $K_T$ extreme line corresponds to $p$-edges as it is the second order derivative of $G$ with respect to $p$, and similarly the $C_p$ extreme line corresponds to $T$-edges.
Combining these, we present the complex $\tilde{p}-\tilde{T}$ phase diagram in Fig.~\ref{fig:vdwCpd}(c), focusing solely on the LY edges.
The unified phase boundary branches beyond the critical point, as different $K_T$ and $C_p$ extreme lines in $x-y$ plane (in Fig.~\ref{fig:vdwCpd}(c)).
Remarkably, these extreme lines largely overlap with the projected trajectories of the LY edges, with small deviation due to contributions from non-edge zeros with large density (inset of Fig.~\ref{fig:vdwCpd}(c)).
It is by retaining complex LY zeros rather than only the real ones that the complex phase diagram embodies the full statistical information.
%

%
One fascinating but intricate fact about the supercritical matter is that there are different boundaries, defined by different thermodynamic properties.
This is intrinsic to the high-dimensional feature of the LY zeros.
When the closest zeros are on the physical plane, this point overrides the others and all properties show maxima at the same place.
But when zeros are away from the physical plane, there can be different edges emanating from the same cluster of zeros corresponding to different physical properties.
As a result, the corresponding extreme lines appear at different places.
These exactly describe the behavior of LY edges and response functions in Fig.~\ref{fig:vdwCpd}(c), where the seemingly two edges curves are 3D projections from a unified 4D zero cluster.
The supercritical region is no longer a ``no transition's land" since we can see how the complex LY edges determine the phase diagram.
In this view, one can interpret supercriticality as a phase transition in the complex phase diagram and an incipient one in the physical plane.

\subsection{2D Ising model}

\begin{figure}[b]
    \centering
    \includegraphics[width=0.95\linewidth]{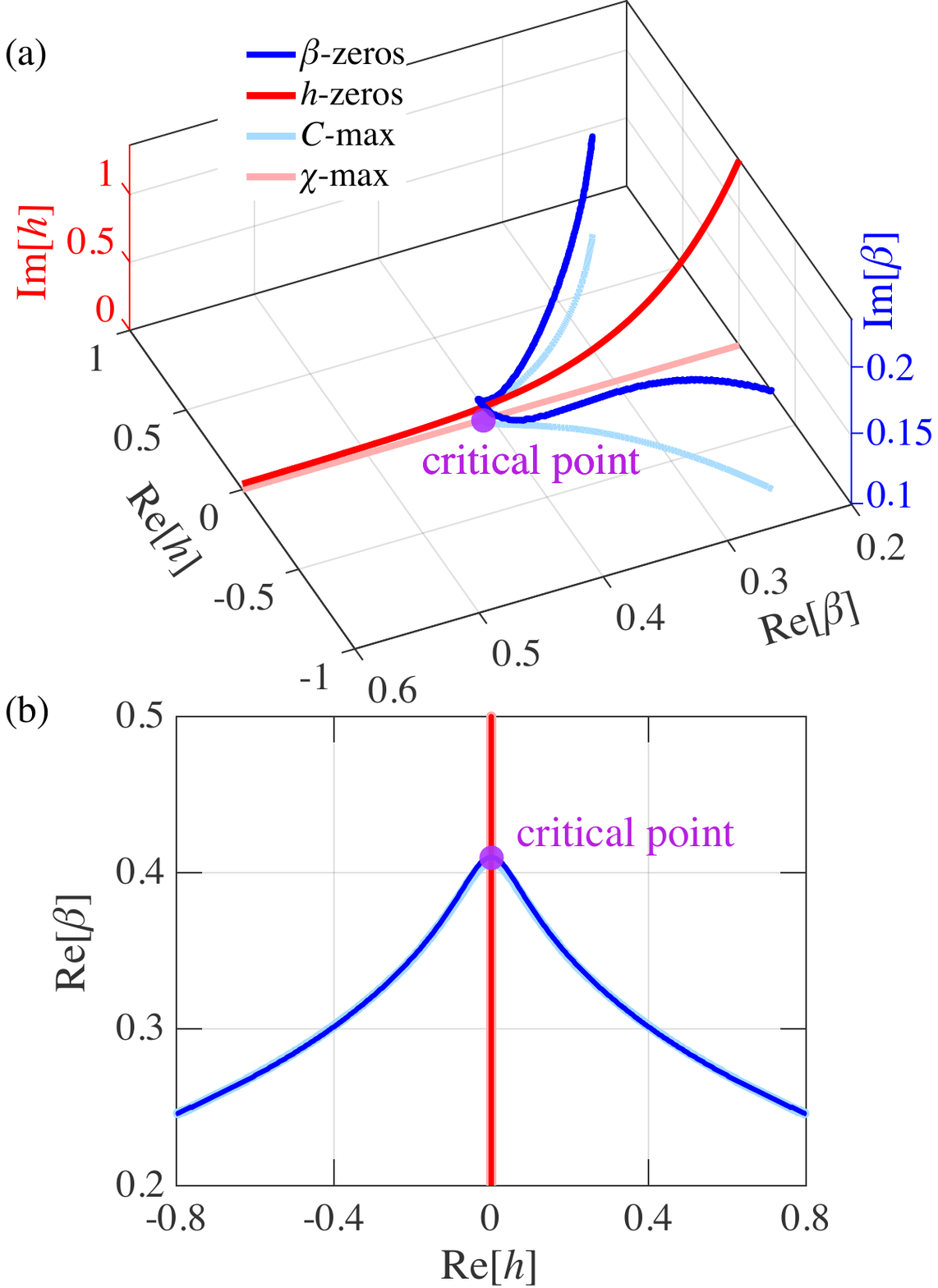}
    \caption{(a) The complex $\tilde{\beta}$-$\tilde{h}$ phase diagram of 2D Ising model, where $\beta=1/T$. (b) The projection of edge zeros to the real plane. The specific heat $C$ almost coincides with the trajectory of $\beta$-edges, so does the susceptibility $\chi$ with $p$-edges.
}
    \label{fig:isingCpd}
\end{figure}

Here, the complex phase diagram of an $8 \times 8$ 2D ferromagnetic Ising model is demonstrated in Fig.~\ref{fig:isingCpd}.
Except for almost coincident phenomena, it is worth noting that the coexistence line in the phase diagram of the Ising model is parallel to the $\beta$-axis \cite{luoBehaviorWidomLine2014}.
Consequently, there are two extreme lines of heat capacity $C$ originated from the critical point, which conforms to the symmetry of system on external magnetic field $h$.
Both of the branches correspond well with the projection of the $\beta$-zeros onto the real plane.
This further corroborate our findings in vdW model.

\subsection{TIP4P water}

\begin{figure}
    \centering
    \includegraphics[width=0.95\linewidth]{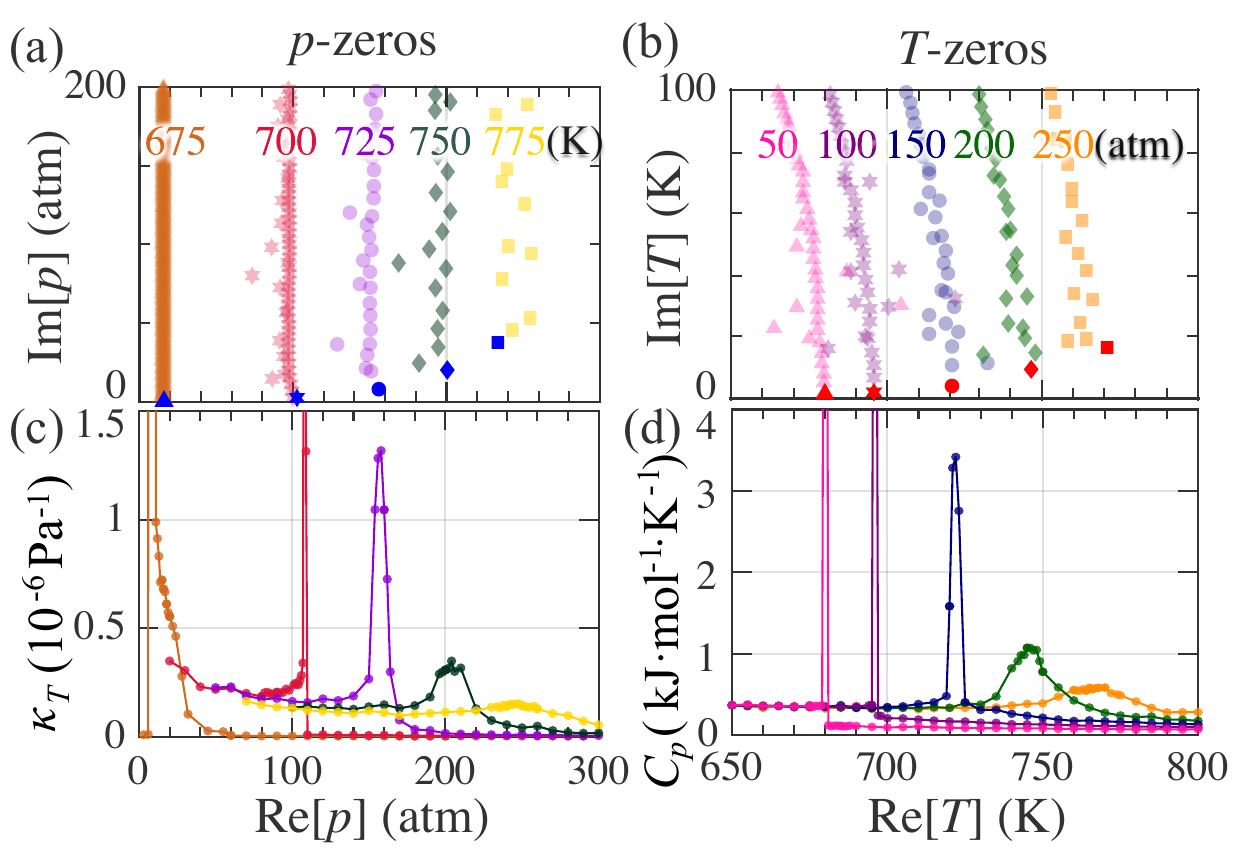}
    \caption{LY zeros and edges calculated from MD simulations of the TIP4P water.
        From left to right: the triangle, star, circular point, diamond, and square marks correspond to
        (a) complex $p$-zeros at $T=675$ to $775~\mathrm{K}$, with blue marks indicating the $p$-edge of each $T$.
        (b) complex $T$-zeros at $p=50$ to $250~\mathrm{atm}$, with red marks indicating the $T$-edge of each $p$.
        (c) Plot of the corresponding isothermal compressibility coefficient $\kappa_T$ with pressure, at temperatures  $T=675$ to $775~\mathrm{K}$.
        (d) Corresponding isobaric heat capacity $C_p$ with temperature, at pressures $p=50$ to $250~\mathrm{atm}$.
        The location $\kappa_T$ or $C_p$ reach its maximum is also close to the corresponding LY edge.
    }
    \label{fig:waterZero}
\end{figure}

\begin{figure}[t]
    \centering
    \includegraphics[width=0.95\linewidth]{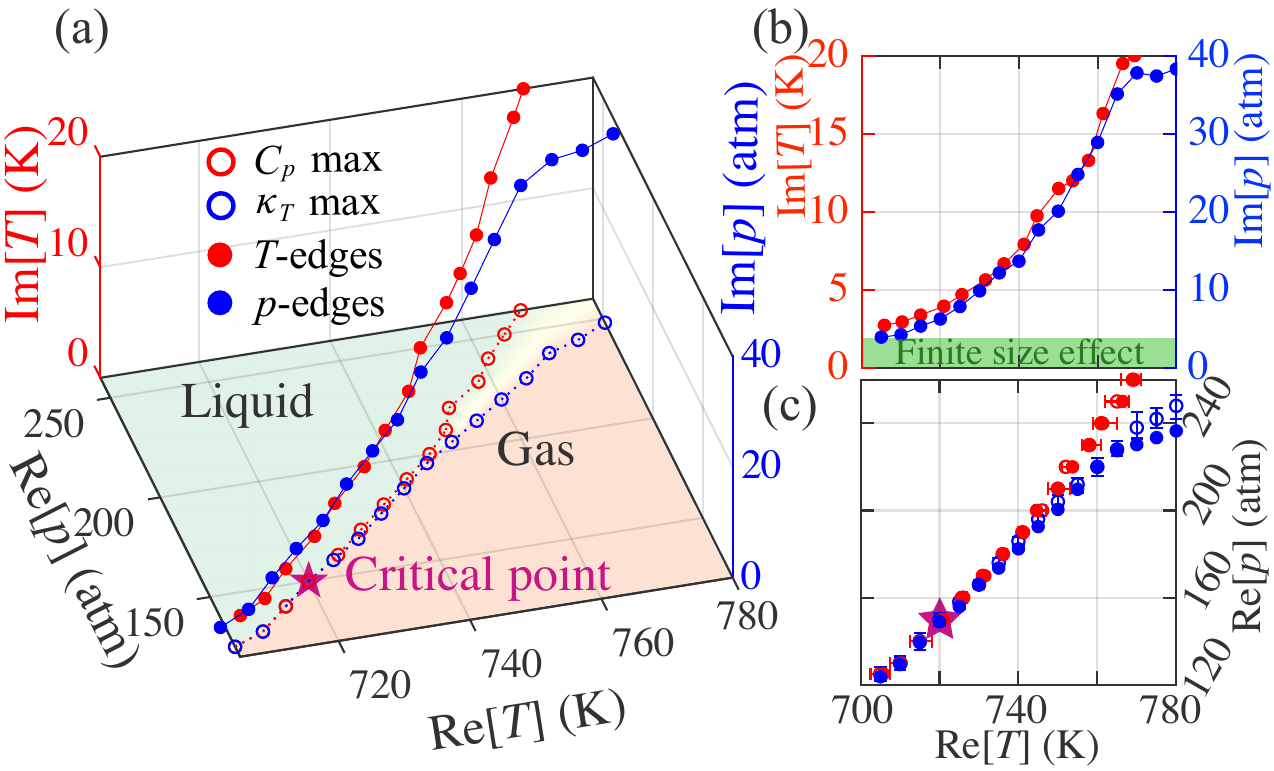}
    \caption{Complex phase diagram of water using the TIP4P model.
        (a) The complex $\tilde{T}$-$\tilde{p}$ phase diagram, where the critical point is determined to be approximately at $T_c\sim720~\mathrm{K}$ and $p_C\sim 150~\mathrm{atm}$.
        (b) The side view and (c) the top view of this complex phase diagram.
        Due to the finite size effect, LY zeros below the critical point approach real axis instead of being exactly onto them.
        Throughout, the Lee-Yang edges corresponding to $T$ and $p$ are labeled with solid marks in red and blue, while the maximum of $C_p$ and $\kappa_T$ are labeled with hollow marks in red and blue, respectively.
    }
    \label{fig:waterCpd}
\end{figure}

Apart from the idealized vdW and Ising models, these traits of the complex phase diagram have also been observed in molecular dynamics simulations of realistic water systems.
Using the method for calculating LY zeros based on MD results, we have derived the $p$- and $T$-zeros (Fig.~\ref{fig:waterZero}(a)(b)) of TIP4P water.
In the meantime, the location $\kappa_T$ or $C_p$ reach its maximum is also close to the corresponding LY edge (Fig.~\ref{fig:waterZero}(c)(d)).
The complex phase diagram of water is given in Fig.~\ref{fig:waterCpd}(a).
It is evident that the $T$-/$p$-edges branch from the coincident transition points and gradually move away from the physical plane (Fig.~\ref{fig:waterCpd}(a)), and manifest similar correspondences to extreme lines with vdW results (Fig.~\ref{fig:waterCpd}(c)).
The consistency in findings across different models and treatments underscores the efficacy of the complex phase diagram.

%
The simulating system size of 216 molecules might induce a finite size effect.
Due to this, zeros cannot approach the real axis within the critical region (Fig.~\ref{fig:waterCpd}(b)).
Besides, the simulation of 4,096 molecules gives critical point $p_c=145~\mathrm{atm},~T_c=640~\mathrm{K}$ \cite{galloWidomLineDynamical2014}, about $80~\mathrm{K}$ larger than our results.
We note elaborate works with considering larger size \cite{galloWidomLineDynamical2014} and other types of transition like liquid-liquid phase transition (LLPT) \cite{xuRelationWidomLine2005} might improve the results quantitatively, however, the conclusions should not be affected.

\section{Discussions}
\subsection{Complex phase diagram: determination of phases by zero structure}
\begin{figure}[b]
    \centering
    \includegraphics[width=0.95\linewidth]{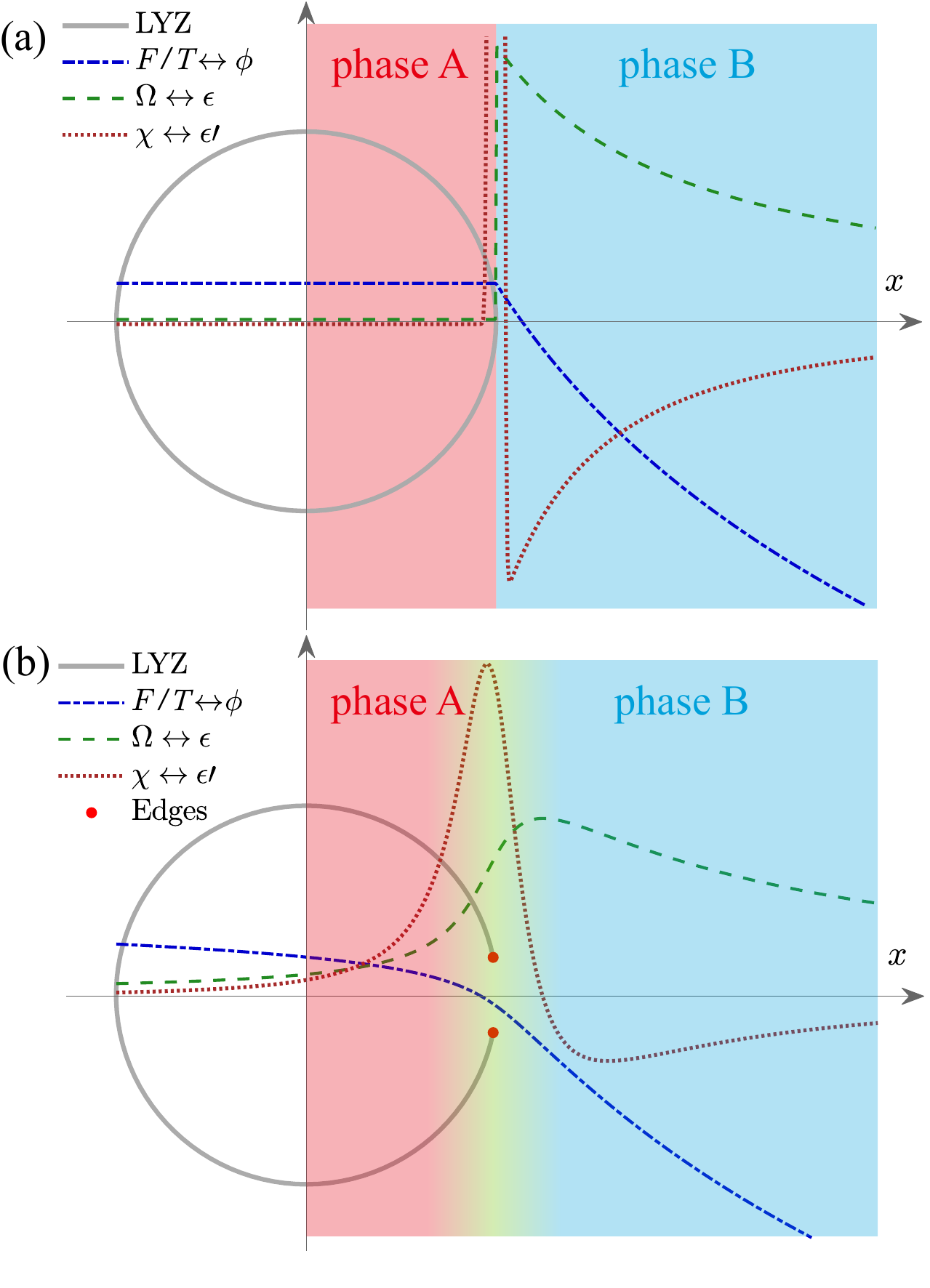}
    \caption{
         A pair of schematics of the electrostatic analogies of LY zeros, to (a) phase transition and (b) crossover.
        (a) Below the critical point, LY zeros (analogous to charges, both in grey solid line) could distribute uniformly on the unit circle.
        This keeps the energetic state function $F$ (analogous to the electric potential $\phi$, both in blue dashed line) a constant inside it and drives $F$ (or $\phi$) decrease outside it, accompanied by a sharp change at the intersection point in order parameter $\Omega$ and susceptibility $\chi$ (analogous to electric field $\epsilon$ and field gradient $\epsilon'$, in green and brown dashed lines, respectively).
        While (b) in the supercritical region, the shell is cut and LY zeros or charges terminate at the edges (red points).
        $F$, $\Omega$, and $\chi$ ($\phi$, $\epsilon$, and $\epsilon'$) manifest continuous changes and finite maxima in the vicinity of the edges.
        The different tendencies on both sides of the edges indicate a crossover of the original two phases.
    }
    \label{Scheme}
\end{figure}

The complex phase diagram employed should evoke a revisit for the definition of ``phase''.
Historically, phases are defined in the viewpoint of phase transition: finding a physical path of transition from one to another, i.e. two phases are distinguished only when abrupt changes occur as real thermal fields vary.
While in the viewpoint of LY zeros, ``a phase'' means unique analytic behaviors within a potential produced by zeros, where the geometric relationship between the location of the system's state and the cluster of zeros dominates.
While these two perspectives converge when there are real zeros, the traditional one falls short in the supercritical region.
Here, crossover replaces phase transition, leading to critical anomalies and inconsistent extreme lines instead of singularities and consistent phase boundaries.
The LY zeros perspective, however, remains robust by providing zero determined complex phase diagram as a unified picture underlying phase transition and crossover.
To elucidate this, we use the electrostatic analogy proposed by Lee and Yang.
Taking the logarithm of Eq.~(\ref{partition function}) and replacing the summation over the discrete zeros with the integral of the (DOZ) $\rho(\tilde{x})$, the energetic state function can be written as~\footnote{Here, the contribution from the exponential factor, which is free of zeros, is continuous and hence be ignored when tackling supercritical anomalies. We note these terms become significant when zeros are distant from the real plane, resulting in the termination of extreme lines.},
\begin{equation}
    {F}(x)\sim -\frac{1}{\beta}\ln Z(x)\approx -\frac{1}{\beta}\int_{\mathcal{C}} \rho(\tilde{x}) \ln (x-\tilde{x})\mathrm{d}\tilde{x},
\end{equation}
where $x$ can be either $T$ or $p$.
This expression is exactly the form of a 2-D Coulomb potential $\phi$ produced by a circular cylinder with surface charge density $\rho(x)$ per unit area.
It means the behavior of order parameter $\Omega$ and susceptibility $\chi$ of the response function can be perceived equivalently from electric field $\epsilon$ and its gradient $\epsilon'$.
The cases of phase transition and crossover are intuitively the analogies of fully screening potential with a closed shell of zeros (Fig.~\ref{Scheme}(a)) and a flux leakage with cuts
in this shell (Fig.~\ref{Scheme}(b)), respectively.
Crossover is the consequence that the field produced by zeros leaks from one phase to another, with strength determined by the cut size (the closest distance of zeros to the physical plane) and the distance to the cut.
%

\subsection{Physical accessibility of zeros}
Considering the purely mathematical origin of zeros, one might wonder if the complex fields corresponding to LY zeros and the complex diagram are physically accessible.
We note that the LY zeros can be obtained either numerically or experimentally~\cite{binekDensityZerosLeeYang1998,weiLeeYangZerosCritical2012,pengExperimentalObservationLeeYang2015,flindtTrajectoryPhaseTransitions2013,brandnerExperimentalDeterminationDynamical2017,brangeLeeYangTheoryBoseEinstein2023,degerLeeYangZerosLargedeviation2018,francisManybodyThermodynamicsQuantum2021,flaschnerObservationDynamicalVortices2018}.
There are primarily two categories of methods to detect LY zeros: (1) the direct detection of zeros, (2) inferring zeros through post-processing of original experimental or simulation data.

The protocol for direct detection was firstly proposed in Ref.~\cite{weiLeeYangZerosCritical2012}, and was initially realized in Ref.~\cite{pengExperimentalObservationLeeYang2015}.
This is done by measuring the quantum coherence of a probe spin coupled to an Ising-type bath, where the evolution of the former relates to complex LY zeros~\cite{weiLeeYangZerosCritical2012,pengExperimentalObservationLeeYang2015}.
This method was later combined with universal quantum computers to overcome numerical difficulties in classical computations, enabling zero detection in a scalable manner as hardware improves~\cite{francisManybodyThermodynamicsQuantum2021}.
Not only for LY zeros in spin systems but also Fisher zeros corresponding to complex $\tilde{T}$ were observed as dynamical vortices involving fermionic atoms in a driven optical lattice~\cite{flaschnerObservationDynamicalVortices2018,brandnerExperimentalDeterminationDynamical2017}.

It would be easier to inferring zeros mathematically.
The closest few zeros to the real axis can be extracted via the high-order cumulant method, which is accessible in both molecular simulations and experiments~\cite{flindtTrajectoryPhaseTransitions2013,degerLeeYangZerosLargedeviation2018,brangeLeeYangTheoryBoseEinstein2023}.
This method is often used to identify phase transition in finite system.
It was firstly proposed in Ref.~\cite{flindtTrajectoryPhaseTransitions2013} to tackle trajectory phase transition in glass models, and later to other phase transitions, including experimental study of quenched phase transition in Ref.~\cite{brandnerExperimentalDeterminationDynamical2017}, molecular zipper in Ref.~\cite{degerLeeYangZerosLargedeviation2018}, and Bose-Einstein condensation in Ref.~\cite{brangeLeeYangTheoryBoseEinstein2023}.
In the meantime, full zeros can be derived from factorizing partition function, albeit with reduced accuracy.
For the first time, Binek derive zeros via analyzing isothermal magnetization data of FeCl$_{2}$ in Ref.~\cite{binekDensityZerosLeeYang1998}.
As shown in Sec.~\ref{sec: discrete zeros}, we discretized the density of states and obtained an approximated polynomial as the partition function.
The results could be improved by combining better sampling methods for density of states.

Besides these methods, we look forward to new experimental techniques to detect the complex phase diagram.

\subsection{Implications of complex phase diagram}
Complex fields can also reveal extra degrees of freedom within the scope of several emerging phenomena, such as the dynamical quantum phase
transition (DQPT), non-Hermitian physics, and non-equilibrium statistics~\cite{heylDynamicalQuantumPhase2013,heylDynamicalQuantumPhase2018,yamamotoTheoryNonHermitianFermionic2019,ashidaNonHermitianPhysics2020,liYangLeeSingularityBCS2022,matsumotoEmbeddingYangLeeQuantum2022,novaMetastableFerroelectricityOptically2019,liTerahertzFieldInduced2019}.
For example, Heyl \textit{et al.} suggested a connection between the thermodynamic phase transition and real-time evolution problems by introducing a complex effective temperature
as $\beta ~\sim \text{i}t$, revealing DQPT as the non-analytical behavior at temporal zeros $t^*$ after quench~\cite{heylDynamicalQuantumPhase2013,heylDynamicalQuantumPhase2018}.
The complex interactions as the coupling of the complex intensive field and real extensive quantities also indicate the non-Hermitian nature of open quantum systems~\cite{yamamotoTheoryNonHermitianFermionic2019,ashidaNonHermitianPhysics2020, liYangLeeSingularityBCS2022,matsumotoEmbeddingYangLeeQuantum2022}.

Concerning metastability, Langer developed a theory using the analytical continuation of the free energy~\cite{langerTheoryCondensationPoint1967,langerStatisticalTheoryDecay1969,guntherNumericalTransfermatrixStudy1993,guntherApplicationConstrainedtransfermatrixMethod1994}.
Contrast to conventional scenario, complex free energy is required in his theory.
We note this would be an immediate conclusion of the complex phase diagram: the free energy can be complex for thermal configurations assigning complex fields.
Besides, the well-known interpretation of imaginary part of energy as lifetime also applies for metastability.
The evolution of a non-Hermitian system, i.e., complex Hamiltonian $H=\Re[H] + i\Im[H]$, writes
\begin{equation}
    e^{iHt}=e^{-\Im[H]t} e^{i\Re[H]t},
\end{equation}
where the decay factor explicitly depends on the imaginary part $\Im[H]$.

Not only metastable but also other nonequilibrium states might find its position in our complex phase diagram, while they can never be accessed in the real phase diagram.
We referred to the experiments which detected “hidden phase” using infrared pulses or terahertz fields in our discussion~\cite{liTerahertzFieldInduced2019,novaMetastableFerroelectricityOptically2019}.
In these experiments, the typically forbidden ferroelectric phase in strontium titanate can be transiently induced by infrared pulses or terahertz fields.
Here, the oscillating field drives the system out of equilibrium and hence out of real phase diagram.
In another example, metamagnetic anomalies might also be interpreted as accumulated results that oscillating fields access configurations of complex phase diagram~\cite{buendiaFluctuationsModelFerromagnetic2017,riegoMetamagneticAnomaliesDynamic2017}.
We anticipate a unified picture containing both equilibrium and nonequilibrium phenomenon stands on the complex phase diagram.

%

\subsection{The $p$-$V$ phase diagram and hidden configurations}
\begin{figure}[htbp]
    \centering
    \includegraphics[width=1\linewidth]{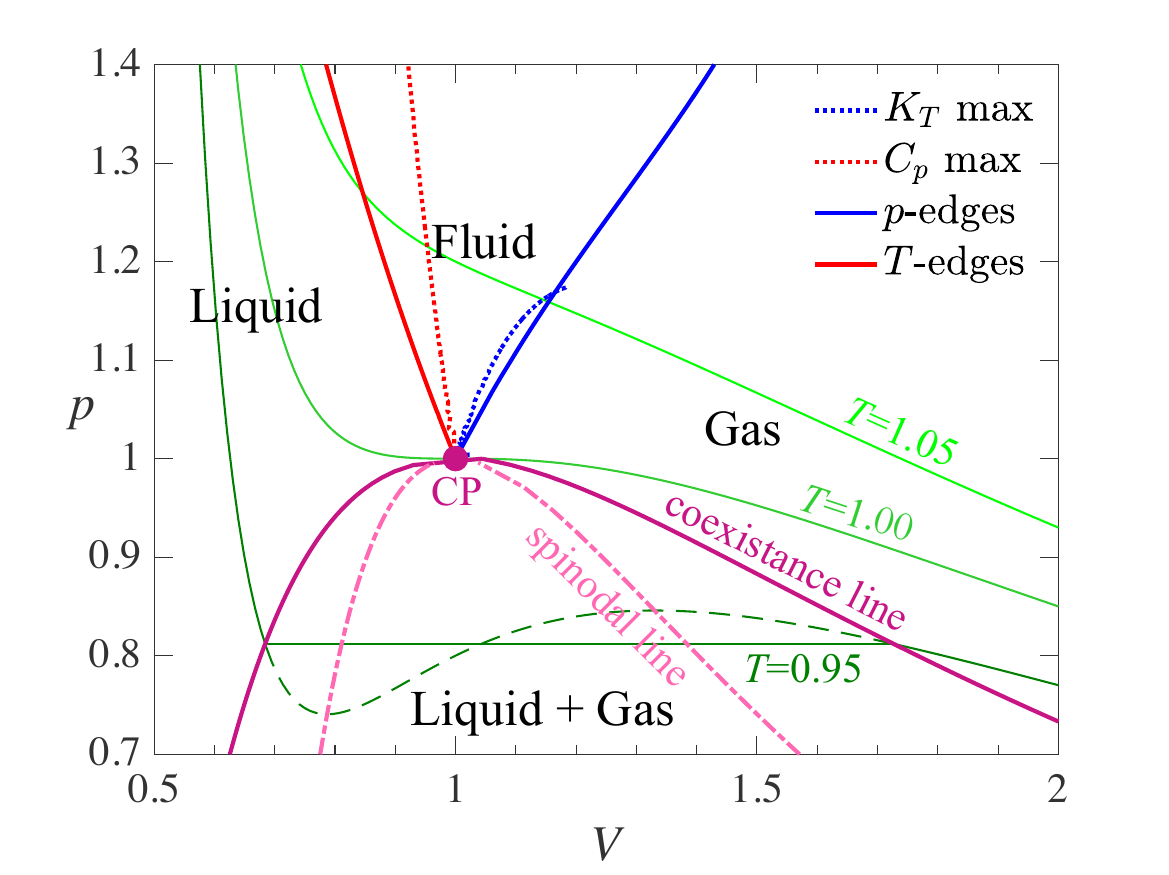}
    \caption{The $p$-$V$ phase diagram of vdW fluids in the vicinity of the critical point. Three isotherms are plotted in green lines, and the coexistence line is plotted with Maxwell's construction. Inside the coexistence line, there is a mixture state of liquid and gas, where metastable states exist outside the spinodal lines. Outside the coexistence line, there is a continuous crossover from liquid-like to gas-like. The $\kappa_T$ and $C_p$ extreme lines and the projections of the $p$- and $T$-edges to the physical $p$-$V$ plane are plotted and used as a boundary of the fluid region. }
    \label{fig:vdwPV}
\end{figure}

In studies of the phase diagram, the choice of the axes favors thermal fields, such as $T$ and $p$.
One of the main reasons is that these fields are external variables, which are independent of the observed system.
Such treatment is associated with constant external field ensembles, such as the isothermal and isobaric ones.
However, there are hidden configurations of the system which is not well-defined and hence cannot be accessed by fixing external fields.
For example, the coexisting state is hidden in the $T$-$p$ phase diagram.
While most $(T,p)$ configurations of a van der Waals (vdW) fluid correspond to a certain volume $V$, there exists a coexistence region below the critical point where $V$ abruptly changes at transition pressure, i.e. a single $(T,p)$ configuration corresponds to a set of states with different $V$s, as shown in Fig.~\ref{fig:vdwPV}.
This implies the limited power of using the conventional real $T$-$p$ phase diagram.

However, as discussed above, the complex phase diagram might be able to describe metastable and nonequilibrium phases and hence includes these coexistence configurations.
From the left branch to the right branch of coexistence line, the configurations conforming to vdW equation of state are known as superheated liquid, unstable one inside spinodal line, supercooled vapor.
It will be interesting to discover them using the complex phase diagram while deeper insights await future studies.

\section{Conclusion}
The LY theory offers fundamental insights into phase transition, particularly highlighting the complex characteristic of zeros.
But its theoretical value is sometimes underestimated due to concerns regarding that zeros have to approach a real axis for a phase transition to occur, or that the complex character of zeros hinders a closer look and more accessible reaches.
In this manuscript, we present a concept of complex phase diagram of higher dimensions than the conventional one used, which contains full information of $T$- and $p$-zeros.
Our work emphasizes the intrinsic role of complex zeros in determining observable phenomena in the real plane, e.g., the supercritical anomalies of thermal responses in $C_p$ and $K_T$ (or $\kappa_T$), and the different supercritical boundaries.
With these, we conclude by saying that the complex zeros stand firmly in physics which merit further experimental explorations with the state of the art of measuring techniques.

\section*{acknowledgments}
The authors acknowledge very insightful discussions with Prof. H. T. Quan and Prof. L. M. Xu. We are supported by the National Natural Science Foundation of China (Grant Nos. 12204015, 12234001, and  11934003), the National Basic Research Programs of China (Grant Nos. 2021YFA1400503 and 2022YFA1403500), National College Students' innovation and entrepreneurship training program (Grant No. 20220064), Beijing Natural Science Foundation (Grant No. Z200004), and the Strategic Priority Research Program of the Chinese Academy of Sciences (Grant No. XDB33010400). The computational resources were provided by the supercomputer center at Peking University, China.

\appendix
\section{requirements for the factorization of partition function}
\label{app: factorization}
In the original paper, Lee and Yang factorize the grand partition function of the 2D lattice gas model (or equivalently the 2D Ising model) as a product of zeros terms of chemical potential (magnetic field).
Since the number of atoms $N$ is always an integer, the partition function as $\exp[-\beta\mu\cdot N]$ is exactly a polynomial of $y=\exp[-\beta\mu]$ with complex zeros $\tilde{y}^*=\exp[-\beta\tilde{\mu}^*]$, where $\beta\mu$ is the chemical potential.
According to the fundamental theorem of algebra, this polynomial is well-established to factorize as the acknowledged form.
Fisher suggested a similar treatment to temperature $T$, known as ``Fisher zeros'' $\widetilde{T}^*$, without proof.
However, it is nontrivial to directly extend LY's idea to arbitrary intensive fields.
In general, their conjugated extensive quantities are not integers, i.e., \textit{for Fisher zeros the extensive quantity energy is real in comparison with the LY zeros where the extensive quantity $N$ is an integer and hence the partition function is no longer a polynomial}.
We note that the partition function is required to be an entire function so that Weierstrass's factorization theorem can be used to factorizing generalized zeros including Fisher zeros.
This is valid when system fulfill certain though physical prerequisites, as shown below.
We consider a general system of monoatomic gas with number of atoms $N$.
To describe its microstates and thermodynamic properties, we use an extensive quantity $A_N = A(N)$ and its conjugated field $\phi_A$.
Using the ensemble theory, the partition function $Z$ is simply the summation of the exponential factors over all possible microstates, as
\begin{equation}
Z_N(\phi_A)=\sum_{\text{all possible microstates}} e^{-\beta A_N\cdot \phi_A}.
\label{eq:partition A}
\end{equation}
For convenience, we use $\beta =1$ in the following.
Introducing the complex field $\tilde{\phi}_A = \phi_A + i\varphi_A$ (or simply $\tilde{\phi}=\phi+i\varphi$), the complex partition function is constructed, as
\begin{equation}
Z_N(\tilde{\phi})=\sum_{\text{all possible microstates}} e^{- A_N\cdot \tilde{\phi}}.
\label{eq:partition A complex}
\end{equation}
The extensive quantity should be proportional to the system size, as $A_{\lambda N}\sim\lambda A_{N},~\lambda\in Z^+$ when the system is enlarged $\lambda$ times.
In fact, the interfacial terms or long-range interactions for non-periodic system would bring deviations from this.
Here, we mainly concern the bulk terms in the thermodynamic limit or suppose the periodic boundary condition.
The following assumptions are made about the nature of the system:
\begin{enumerate}
    \item The particles have hard cores so that a system with finite size can contain only finite atoms. Or simply, the density of particles is finite in the thermodynamic limit. This is also assumed by Lee and Yang in Ref.~\cite{yangStatisticalTheoryEquations1952}.
    \item The normalized density of states (or probability distribution density of states) $p(A_N)$ decays much faster than $e^{-N}$. Lee and yang assumed the interaction has a finite range so that they need not consider infinite $A$. Here, we generalize this.
    \item For any $N$, the averaged contribution of a single atom to the extensive quantity is nowhere negatively infinite, as $\lim_{N\to\infty}\{A_N/N\}=a\neq-\infty$. The assumption that $u(r)$ is nowhere minus infinity in Ref.~\cite{yangStatisticalTheoryEquations1952} is a special case for energy.
\end{enumerate}

The second assumption can in principle be interpreted that there should be no infinite characteristic scale in physical system.
On one hand, $p_N(A\to\infty) \to 0$ otherwise the infinite $A$ contributes significantly.
On the other hand, if a system of scale $N_0$ contains all its physics, then the density of states of an enlarged system can be viewed as multiple replicas of scale $N_0$.
The normalized density of states is given by convolution products, as
\begin{equation}
  p(A_{\lambda N_0}) = \sum_{\sum_{i=1}^{\lambda}A_{N_0,i}=A_{\lambda N_0}} \prod_{i=1}^{\lambda} p(A_{N_0,i}).
\end{equation}
Consequently, \begin{equation}
  \lim_{\lambda\to\infty}|p(A_{\lambda N_0})|^{1/\lambda}\to 0.
\end{equation}

According to the third assumption, $\{A_n/n\}$ where $n=1,\cdots,\infty$ is a bounded sequence.
Consequently, it has a lower boundary $a_0$ that $A_n/n\ge a_0,~\forall n$.
Whilst the upper boundary seems unnecessary due to the negative exponential factor and hence insignificant statistical contribution.
It should be noted that this condition is associated with the first assumption.
If it is not true, e.g. interactions between particles can be negatively infinite, then more and more even infinite particles will be attracted to the system.
This is contradictory to the restriction of hard cores.

Considering the distribution of $A$, the partition function in Eq.~(\ref{eq:partition A complex}) is written as
\begin{equation}
    Z_N(\tilde{\phi}) = \sum_{k=1}^{\infty} n_{N,k} \exp[-\tilde{\phi} \cdot A_{N,k}],
    \label{discrete form}
\end{equation}
when $A_N$ takes discrete values $A_{N,k}$, where $\forall i<j,~Na_0 < A_{N,i} < A_{N,j}$, and by
\begin{equation}
    Z_N(\tilde{\phi}) = \int_{N a_0}^{\infty} \rho_N(A_N) e^{-\tilde{\phi}\cdot A_N} dA_N,
    \label{continuous_form}
\end{equation}
when $A$ distributes continuously.
Here, $n_{N,k}$ and $\rho_N(A_N)$ are the degeneracy of state and the density of state, respectively.
Noting that a constant is reducible in ensemble statistics, we shall normalize them as the probability distribution function $p_{N,k}$ or $p_{N}(A_N)$, which satisfy $\sum_{k} p_{N,k} = 1$ or $\int_{N a_0}^{\infty} p_{N}(A_N) dA_N =1$, respectively.
To demonstrate that $Z_N(\tilde{\phi})$ is an entire function is equivalent to check if $Z_N(\tilde{\phi})$ is holomorphic with arbitrary $\tilde{\phi}$ value.
We shall show the proof for the discrete case in Eq.~(\ref{discrete form}).
The continuous case for Eq.~(\ref{continuous_form}) has a similar logic.
Firstly, let us consider the case when the system is of finite size $N$.
When there are finite levels of $A_k$, $Z_N(\tilde{\phi})$ is a finite sum of exponential functions and is obviously an entire function.
Otherwise, there are infinite levels of $A_k$.
Note that the exponential factor is always positive and decreases with increasing $A_{N,k}$, the absolute summation of terms in $Z_N(\tilde{\phi})$ equals $|Z_N(\tilde{\phi})|$, satisfying
\begin{equation}
    \begin{split}
        |Z_N(\tilde{\phi})| &= \sum_{k=1}^{\infty} \left|p_{N,k} \exp[-|\tilde{\phi} \cdot A_{N,k}]\right| \\
        &= \sum_{k=1}^{\infty} p_{N,k} \exp[-\Re[\tilde{\phi}] \cdot A_{N,k}]\\
        &\le \sum_{k=1}^{\infty} p_{N,k} \exp[-\Re[\tilde{\phi}] \cdot A_{N,1}]\\
        &= \exp[-\Re[\tilde{\phi}] \cdot A_{N,1}] =|\exp[-\tilde{\phi}\cdot A_{N,1}]|.
    \end{split}
\end{equation}
Since $\exp[-\tilde{\phi} \cdot A_{N,1}]$ is a typical entire function whose radius of convergence is infinite, $Z_N(\tilde{\phi})$ also converges absolutely according to the comparison test.
Thus, for finite $N$, $Z_N(\tilde{\phi})$ is holomorphic in the whole complex plane and is also an entire function.
Then we consider the case in the thermodynamic limit $N\to\infty$.
Note that Eq.~(\ref{discrete form}) can be rewritten in argument $y = \exp[-\tilde{\phi}]$, as
\begin{equation}
    \begin{split}
        \lim_{N\to\infty} Z_N(\tilde{\phi}) &= \lim_{N\to\infty} \sum_{k=1}^{\infty} p_{N,k} y^{A_{N,k}}\\
        &=\lim_{\lambda\to\infty} \sum_{k=1}^{\infty} p_{\lambda N_0,k} y^{A_{\lambda N_0,k}} \\
        & \sim \lim_{\lambda\to\infty} \sum_{k=1}^{\infty} p_{\lambda N_0,k} y^{\lambda A_{N_0,k}}.
    \end{split}
\end{equation}
It turns to prove the polynomial-like series converges in the complex plane.
According to preliminary assumptions, this is true since the coefficient and the index satisfies
\begin{equation}
    \lim_{\lambda \to \infty} |p_{\lambda N_0,k}|^{\frac{1}{\lambda A_{N_0,k}}} = \lim_{\lambda \to \infty} \left\{|p_{\lambda N_0,k}|^{\frac{1}{\lambda}}\right\}^{1/A_{N_0,k}} = 0.
\end{equation}
Combining these, the partition function is entire both for finite system size and in the thermodynamic limit.
Finite lower bound for the corresponding extensive quantity and fast decayed density of states are important for this to hold true, which is always fulfilled for the realistic system.

Thus, the partition function $Z_N(\tilde{\phi})$ of almost all ensembles can be represented as a possibly infinite product involving its zeroes.
Weierstrass's factorization theorem claims the existence of an entire function $g$ and a sequence of integers $\{p_i\}$ such that $Z_N(A)$ can be factorized, as
\begin{equation}
    Z(\tilde{\phi}) = \tilde{\phi}^m e^{g(\tilde{\phi})} \prod_{k=1}^{\infty} E_{p_i}\left(\frac{\tilde{\phi}}{\psi_i^*}\right),
\end{equation}
where $\psi_i^*$ are the zeros, $m=0$ since $Z(\tilde{\phi}=0)\neq 0$, and $E_p(z)$ takes the form of
\begin{equation}
    E_p(z) = \left\{ \begin{split}
        &1-z, & p=0\\
        &(1-z) \exp\left[\sum_{k=1}^p \frac{z^k}{k} \right], & p>0,
    \end{split}\right.
\end{equation}
Physical interests are mainly on the zeros, underlying which the critical behaviors and anomalies occur.
Therefore, we extract the non-zero terms including $e^{g(\tilde{\phi})}$ and exponential terms of $E_p(\tilde{\phi}/\psi_i^*)$ into an analytic function and rewrite it in forms of zeros, as
\begin{equation}
    Z(\tilde{\phi}) = e^{h(\tilde{\phi})}  \prod_{k=1}^{\infty} \left(1-\frac{\tilde{\phi}}{\psi_k^*}\right),
    \label{single-field factorization}
\end{equation}
where the form of $e^{h(\tilde{\phi})}$ is used to indicate that this part is analytic and never takes zero values.

\section{factorization within multiple fields}
\label{app: multiple field}
Eq.~(\ref{single-field factorization}) shows the factorization when there is a single field.
However, realistic system is described by multiple fields, such as $T$ and $p$.
$Z(T,p)$ can only be factorized according to $T$ and $p$, respectively.
To the best of our knowledge, there is no factorization theorem for multiple variables.
However, this problem can be convert to the above case.
One can perform factorization by one field variable when the other field variables are fixed.
That is, the multi-variable function $f(x_1, \cdots, x_n)$ becomes a function of $x_1$ when fixing $(x_2=x_2^\prime,x_3=x_3^\prime,\cdots,x_n=x_n^\prime)$, as $F(x_1)|_{x_2^\prime,x_3^\prime,\cdots,x_n^\prime}$.
If $F(x_1)$ takes zero value at $x_1^*$, then $f(x_1, \cdots, x_n)$ takes zero value at $ (x_1^*, x_2^\prime,x_3^\prime,\cdots,x_n^\prime)$, as
\begin{equation}
    \begin{split}
        f(x_1^*, &x_2^\prime,x_3^\prime,\cdots,x_n^\prime) = F(x_1^*)|_{x_2^\prime,x_3^\prime,\cdots,x_n^\prime} = 0.
    \end{split}
\end{equation}
Here, the dependency of $x_1^*$ on $(x_2,\cdots,x_n)$ is expressed by
\begin{equation}
    x_1^* = g(x_2, x_3, \cdots, x_n).
\end{equation}
Therefore, one can factorize $f$ as
\begin{equation}
    f(x_1, x_2,\cdots,x_n) \sim \prod_{g} (x_1 - g(x_2,x_3,\cdots,x_n)).
\end{equation}
For the simplest case of $Z(T,p)$, one can scan the zeros slice by slice, i.e. varying $p$ and derive zeros corresponding to complex $T$ for each fixed $p$, and vise versa.
In so doing, we factorized the partition function by $p$-zeros or $T$-zeros, as Eq.~(\ref{partition function}) in the main text.
The high-dimensional nature of zeros is manifested by assembling zeros in these slices to a unified distribution in complex space of multiple field.

\section{the termination of extreme lines}
\label{App: termination of extreme line}
According to the rigorous factorization formula Eq.~(\ref{single-field factorization}), there are two terms in $Z(\tilde{x})$ which contribute differently to free energy.
Being always nonzero, the exponential part $e^{h(\tilde{x})}$ contributes to the properties of the system but never induces singularity, while the product related to zeros $\prod\left(1-\tilde{x}/x^*\right)$ induces singularity.
For instance, considering a single temperature field $\tilde{x}=T$ and $T$-zeros, the free energy writes
\begin{equation}
    \begin{split}
        F(T) &\sim -\frac{1}{k_B T}\ln [Z(T)] \\
        &= -\frac{1}{k_B T}\left[ h(T) + \sum_l \ln [T - \widetilde{T}^*_l] \right] \\
        &=-\frac{1}{k_B T}\left[ h(T) + \int_{\mathcal{C}} \rho(\widetilde{T}^*) \ln [T - \widetilde{T}^*] d\widetilde{T}^* \right],
    \end{split}
\end{equation}
where the summation over all zeros in the second terms is equivalently rewritten as an integral over the density of zeros.
The specific heat is given by
\begin{equation}
    \begin{split}
        C =& - T \frac{\partial^2 F}{\partial T^2}\\
        =& \frac{2}{k_B T^2} \left[ h(T) + \int_{\mathcal{C}} \rho(\widetilde{T}^*) \ln [T - \widetilde{T}^*] d\widetilde{T}^* \right]\\
        & -  \frac{2}{k_B T} \left[ h^\prime(T) + \int_{\mathcal{C}} \frac{\rho(\widetilde{T}^*)}{T - \widetilde{T}^*} d\widetilde{T}^* \right]\\
        &+  \frac{1}{k_B} \left[ h^{\prime\prime}(T) - \int_{\mathcal{C}} \frac{\rho(\widetilde{T}^*)}{(T - \widetilde{T}^*)^2} d\widetilde{T}^* \right].
    \end{split}
\end{equation}
We ignore the contribution from the continuous part in the main text since we focus on phase transition and crossover region therein.
Especially when the distance between the positions of thermal configurations and zeros $|x-\widetilde{x}^*|$ is small, the thermal properties are mainly determined by the latter term related to zeros.
However, the first term turns more significant for regions far away from phase boundary and critical point, where the distance $|T-\widetilde{T}^*|$ is large.
While the contribution related to the density of zeros would quickly decay with increasing $|T-\widetilde{T}^*|$, the contribution related to $h(x)$ and its derivatives are independent of this.
Consequently, it would be difficult even no longer able to perceive critical anomalies induced by zero-related terms.
This corresponds to the termination of the extreme lines.
In the main text, we have shown the numerical results of vdW and water to verify this theoretical proposal.
As shown in Fig.~\ref{fig:vdwWidom}, the maximum values of $K_T$ cannot be distinguished when $T>1.07$, and the maximum values of $C_p$ gradually become insignificant for higher $p$s too.
The similar conclusion could be found for water in Fig.~\ref{fig:waterWidom}, where the $\kappa_T$ peak become insignificant for $T>800$K.

%

\end{document}